\documentclass[aps,nofootinbib,prr,reprint,superscriptaddress,longbibliography]{revtex4-1}
\usepackage{amsmath}
\usepackage{graphicx}
\usepackage{amsfonts}
\usepackage{color}
\usepackage{bm}
\usepackage{lipsum}
\usepackage{comment}
\usepackage{dsfont}
\usepackage{bm}
\usepackage{amsmath}
\usepackage{amssymb}
\usepackage{graphicx} 
\usepackage{braket}
\usepackage{mathtools}
\usepackage{times}
\usepackage{enumerate}
\usepackage{subcaption}

\newcommand{\ii}{\mathrm{i}}

\newcommand{\bea}{\begin{eqnarray}}
\newcommand{\eea}{\end{eqnarray}}
\usepackage[colorlinks,linkcolor=blue,urlcolor=blue,citecolor=blue]{hyperref}

\begin{document}

\title{Suppression of decoherence dynamics by a dissipative bath at strong coupling}

\author{Jitian Chen$^{\times}$}
\email{jitian.chen@mail.utoronto.ca}
\affiliation{Department of Chemistry, University of Toronto, 80 Saint George St., Toronto, Ontario, M5S 3H6, Canada}

\author{Jakub Garwo\l a$^{\times}$}
\email{jakub.garwola@mail.utoronto.ca}
\affiliation{Department of Physics and Centre for Quantum Information and Quantum Control, University of Toronto, 60 Saint George St., Toronto, Ontario, M5S 1A7, Canada}

\author{Dvira Segal}
\email{dvira.segal@utoronto.ca}
\affiliation{Department of Chemistry, University of Toronto, 80 Saint George St., Toronto, Ontario, M5S 3H6, Canada}
\affiliation{Department of Physics and Centre for Quantum Information and Quantum Control, University of Toronto, 60 Saint George St., Toronto, Ontario, M5S 1A7, Canada}

\def\thefootnote{$\times$}\footnotetext{These authors contributed equally to this work. }

\date{\today}

\begin{abstract}
Control of decoherence in open quantum systems has become a topic of great interest due to the emergence of quantum technologies that depend on quantum coherent effects. 
In this work, we investigate the decoherence dynamics of systems coupled to multiple baths through noncommuting systems' operators, and beyond the weak system-bath coupling limit. By building on cooperative effects between baths, we propose a novel strategy to mitigate rapid decoherence. Concretely, we study the dynamics of a qubit coupled to multiple environments with arbitrary interaction strengths, and along different coordinates. 
Based on insights gained on the decoherence dynamics from the analytical Effective Hamiltonian method, we carry out numerical simulations using the Reaction Coordinate quantum master equation method. In contrast to standard expectations, we show that when the system strongly interacts with a decohering bath, increasing its coupling to a second, dissipative bath {\it slows down} the decoherence dynamics. Our work offers insights into the preservation of quantum coherences in open quantum systems based on frustration effects, by utilizing cooperative effects between different heat baths. 
\end{abstract}

\maketitle

\section{Introduction}
\label{sec:Introduction}

Quantum coherence is an important property of quantum mechanical systems, which describes the ``quantumness" of the system, or the ability for quantum states to interfere, and the system to exhibit nonclassical behaviors such as quantum entanglement \cite{RevC,RevQC}.
Quantum coherences enable for a wide range of applications \cite{RevC,RevQC}: In quantum computing, it is essential to maintain superpositions and entanglement for quantum speedup in algorithms \cite{Qalgo}. In quantum metrology, quantum coherences can enhance measurement precision beyond the classical limit \cite{Qmet}. Quantum communication protocols leverage coherence to achieve secure information transfer \cite{QKD}. Moreover, coherences between quantum states play a fundamental role in other areas, such as coherent control of chemical reactions \cite{BrumerBook} and charge and energy transfer on the molecular scale \cite{Scholes18}. 

Noisy Intermediate-Scale Quantum (NISQ) devices represent the current generation of quantum processors, characterized by a limited number of qubits and the absence of error correction \cite{Preskill18,chen_complexity_2023}. Due to unavoidable crosstalk and interactions with their environments, these devices experience decoherence and relaxation, which degrade their quantum state over time. Two phenomenological parameters quantify these errors: the population relaxation time, $T_{1}$, and the phase decoherence time, $T_2$. The relaxation (or dissipation) time $T_1$ describes the timescale over which a qubit loses energy due to dissipative processes. 
The decoherence time $T_2$ quantifies the loss of phase coherence in a superposition state, arising from both the energy relaxation and the pure dephasing mechanisms. These timescales eventually determine the fidelity of quantum computations and the feasibility of running deep quantum circuits on NISQ hardware \cite{10.1063/1.5088164}. Extending these times is therefore a central challenge in the development of quantum processing units. 

Dynamical decoupling (DD) is a prominent strategy in fighting against decoherence. It involves a sequence of pulses designed to counteract the effect of the environment and essentially decouple the system from it \cite{Lorenza98,Lorenza99,Goetz, DDnoise,Lidar23,motlakunta_preserving_2024,Brown_2015}. Other methods include the Quantum Zeno Effect (QZE), Quantum Error Correction (QEC), and Optimal Control Theory (OCT) \cite{Lidar05,Stock}. QZE relies on frequent measurements to inhibit the evolution of a quantum system, QEC detects and corrects errors through redundancy and feedback mechanisms, and OCT employs numerical optimization to design time-dependent pulse sequences that minimize decoherence. All of these involve time-varying operations to counteract decoherence \cite{home1997conceptual, steane1996error, treutlein2006microwave}. Unlike these and other time-dependent control schemes \cite{Kur04,Lidar05,Stock}, the approach described in this work is autonomous; it does not involve time-dependent fields.

A wealth of theoretical approaches have been developed to describe the effect of energy dissipation and quantum decoherence in open quantum systems \cite{BookOQS,NitzanBook}. In this work, we use the quantum master equation (QME) family of method. The well-known Redfield QME is derived from microscopic principles under the assumption of Markovian dynamics and weak couplings to the baths \cite{NitzanBook}, leading to additive impacts of the baths on the system. However, these limitations can be overcome using Markovian embedding methods: In essence, Markovian embedding works by redefining the system-bath boundary, allowing a Markovian weak-coupling approach to be applied to an extended system.

Since its inception for studying chemical reactions \cite{Irene1, Irene2}, the Reaction Coordinate (RC) method---a Markovian embedding technique---has gained increasing attention as a powerful and easy-to-implement approach for open quantum systems \cite{Nazir2018}. Essentially, this method involves performing simulations using Lindblad or Redfield QMEs within an extended Hilbert space, where the system remains weakly coupled to a residual environment assumed to be Markovian. We refer to this computational framework as RC-QME simulation.
Recently, RC-QME simulations have been applied to various problems, including pure decoherence dynamics \cite{antoD}, quantum dissipation in impurities \cite{Ahsan16} and their steady state \cite{Latune22},
quantum thermal transport \cite{Strasberg2016,AhasanE17, FelixS, FelixQAR,correa19,Nazir24stat,Fran24}, dissipation-engineered magnetic order \cite{BrettPRL}, as well as fermionic transport \cite{StrasbergF18,GernotE19,Gernot21Filter,AhsanTE22}. 
However, while powerful, the RC-QME method primarily yields numerical results rather than analytical insights, and thus does not provide a clear strategy to suppress or eliminate decoherence and relaxation processes.

Complementing numerical techniques such as the RC-QME, a promising analytical approach, the Effective Hamiltonian (EFFH) method, has recently been developed in Ref. \citenum{anto2023effective}. The EFFH method begins by extracting a collective reaction coordinate mode from each harmonic bath, similar to the RC mapping. Then, a polaron transformation is applied, followed by a truncation of the polaron-shifted reaction coordinate modes, thus yielding the so-called Effective Hamiltonian. In this framework, strong coupling effects are encoded in renormalized parameters, those of the system, and of its interaction with the baths.
The EFFH method has been proven to be valuable for studying the dynamics \cite{garwola2024open}, equilibrium properties \cite{BrettEQ,BrettSSH}, and nonequilibrium steady states \cite{anto2023effective} of various open quantum system models with arbitrary coupling strengths. Recently, the EFFH framework was further extended to handle systems coupled to multiple baths through noncommuting system's operators \cite{garwola2024open}.

\begin{figure}[htbp]
\centering
\includegraphics[width=0.49\textwidth]{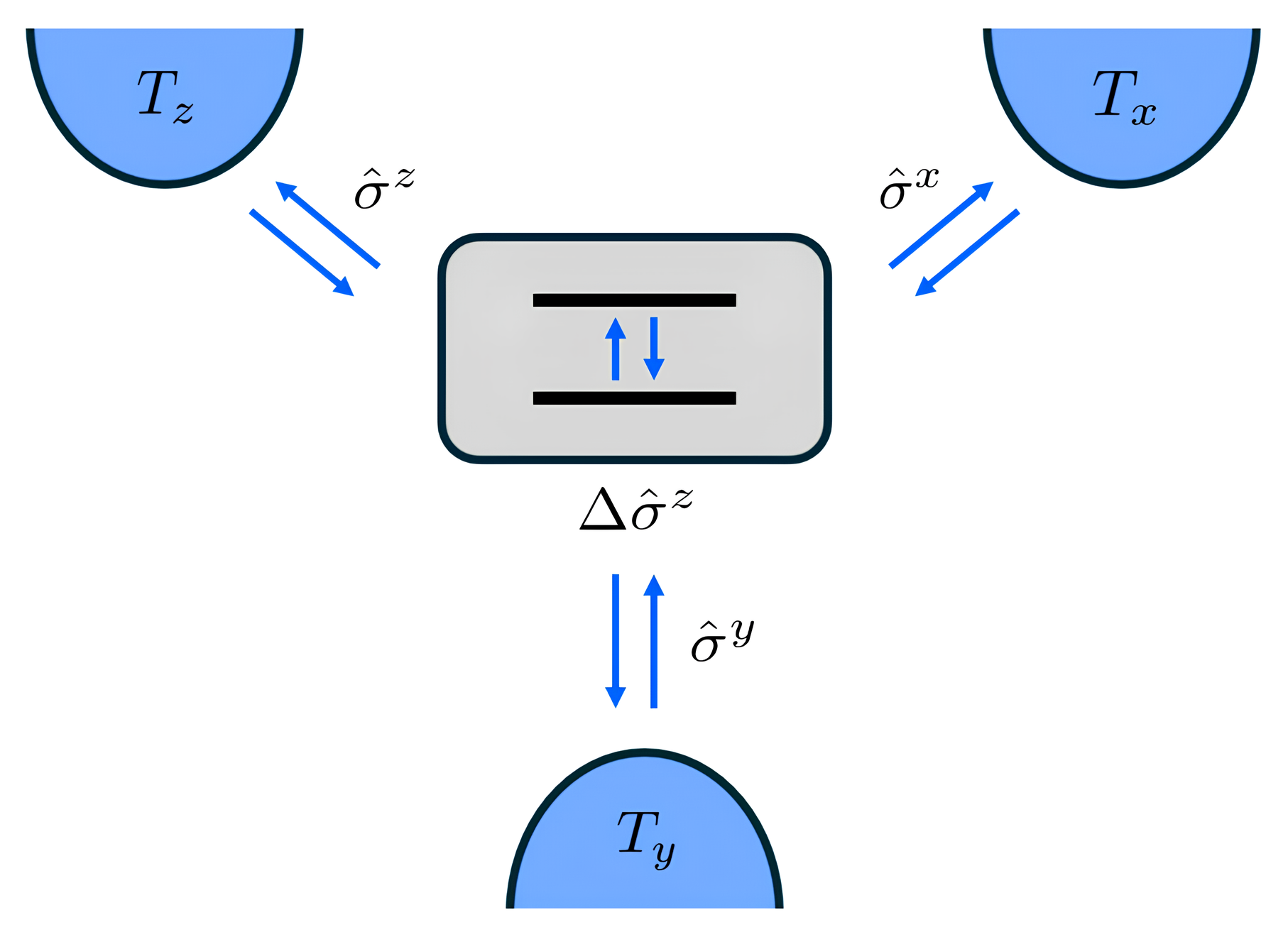} 
\caption{Scheme of our model: A two-level (qubit) system couples to three independent heat baths through noncommuting operators, $\hat\sigma^x$, $\hat\sigma^y$, and $\hat\sigma^z$. $\Delta$ is the spin splitting.
Each bath couples to the system with a different strength, characterized by the interaction parameters $\lambda_{x,y,z}$. For simplicity, we assume that all baths share the same spectral properties with a cutoff frequency $\Omega$ and spectral width $\gamma$. We also set their temperatures to $T=T_x = T_y = T_z$.}
\label{fig:scheme}
\end{figure}

To explore the interplay between decoherence and relaxation, we consider the setup illustrated in Fig. \ref{fig:scheme}, where a qubit is coupled to multiple heat baths through noncommuting system operators. 
The suppression of population relaxation by enhancing the coupling of a qubit to a decohering ($z$) bath was demonstrated numerically in Ref. \cite{Ahsan19,NalbachPRA21}. It was later analytically explained in Ref. \citenum{garwola2024open}. However, in a previous study the decoherence rate was shown to always increase with increasing any of the couplings to the two baths \cite{NalbachPRA21}. In contrast, in this work, we uncover the highly nonintuitive effect of {\it suppressing decoherence} by {\it increasing} the coupling of the system to a dissipative bath. This effect arises from cooperative effects between the baths, leading to ``frustration", which is induced by the noncommutativity of system operators beyond the weak coupling limit. Combining numerical and analytical techniques, we predict the existence of this effect, identify the parameter regimes in which it manifests, and demonstrate it with simulations.

Our main findings are twofold. First, we derive the effective Hamiltonian for the model in Fig. \ref{fig:scheme}, given in Eq. (\ref{eq:HeffG}). Beyond its relevance to study quantum impurity dynamics and their steady-state properties, this effective Hamiltonian can be easily extended to study, for example, the magnetic phases of lattice models coupled to multiple local baths \cite{BrettPRL}. Second, we identify the phenomenon of decoherence suppression via relaxation, demonstrating it through analytical insights (Eq. (\ref{eq:decoh_rate}) and Fig. \ref{fig:Contour}) and supporting it with numerical simulations (Fig. \ref{fig:Control_decay_rate}).

The paper is organized as follows. In Section \ref{sec:Model}, we introduce our model, consisting of a system coupled to multiple baths through noncommuting operators. We discuss our methods in Sec. \ref{sec:QME}: The Redfield QME in Sec. \ref{sec:redfield} and the RC Markovian embedding method in Sec. \ref{sec:strong_coupling}. In Sec. \ref{sec:Effective-Hamiltonian-theory}, we provide a brief overview of the EFFH method, highlighting its mathematical formulation and limitations. 
Predictions based on the EFFH method are included in Sec. \ref{sec:resultsEFFH}. Guided by these results, RC-QME simulations are presented in Section \ref{sec:Results}, focusing on the effect of decoherence suppression by increasing the coupling to a dissipative bath.
We conclude in Section \ref{sec:Conclusions} with a summary of our findings.


\section{Model}
\label{sec:Model}

We consider a single spin (qubit), characterized by the energy splitting $\Delta$ and the tunneling rate $E$. The system Hamiltonian is given by $\hat H_S=\Delta \hat\sigma^z + E \hat \sigma^x$. This qubit is coupled to two or three independent bosonic baths, indexed by $\alpha$. Each bath includes a set of harmonic oscillators, indexed by their momentum $k$. We write the total Hamiltonian in the three-bath case as ($\hbar = 1$),
\bea
&&\hat{H} 
=\Delta\hat{\sigma}^{z}+E\hat{\sigma}^{x}
\nonumber\\
&&+\sum_{\alpha=x,y,z}\sum_k\left[t_{\alpha,k}\hat{\sigma}^{\alpha}(\hat{c}_{\alpha,k}^{\dagger}+\hat{c}_{\alpha,k})+\nu_{\alpha,k}\hat{c}_{\alpha,k}^{\dagger}\hat{c}_{\alpha,k}\right].
\label{eq:H_original}
\eea
Here, $\nu_{\alpha,k}$ and $t_{\alpha,k}$ 
represent the frequency of the $k$th mode of the $\alpha$ bath and its coupling strength to the system. We denote the baths by $\alpha=x,y,z$, identifying along which spin orientation they couple to the system.
The operators $\hat{c}_{\alpha,k}$ are bosonic annihilation operators of the corresponding mode and bath. The system's operators $\hat{\sigma}^\alpha$ couple to the displacements of the baths, and are given in terms of the $\alpha=x,y,z$ Pauli matrices. The coupling to the environment can be described by spectral density functions $J^\alpha(\omega)=\sum_k |t_{\alpha,k}|^2 \delta(\omega-\nu_{\alpha,k})$.

We can construct four nontrivial models, denoted by XY, YZ, XZ and XYZ. We refer to each configuration using its bath indices. 
For example, if the system couples to two baths through its $\{\alpha\}=\{x,y\}$ Pauli components, the model will be referred to as the XY model. 
In the XYZ model, the qubit couples to three baths, one for each of its spin orientation, as recently discussed in Ref. \citenum{AndersQ24}. 

In this work, we focus on the XZ model, which represents coupling to both a dissipative bath ($x$) and a decoherring bath ($z$). We further discuss the Effective Hamiltonian mapping of the XYZ model.
It is important to note that we deliberately choose noncommuting coupling operators to the different baths. We refer to this scenario as ``frustration", since the qubit is coupled through different spin components to the different baths, thus missing a ``preferred" direction. As we show in Sec. \ref{sec:Results}, this aspect is essential to demonstrate suppression of decoherence dynamics. 

\section{Methods}
\label{sec:QME}
We next summarize the methods we use: the Redfield QME, which is valid only in the weak-coupling regime for Markovian baths; the RC-QME, which employs Markovian embedding to capture strong-coupling effects; and the EFFH method, which builds on the RC mapping to provide an approximate yet nonperturbative analytical solution. 

\subsection{Weak coupling: The Redfield QME}
\label{sec:redfield}

We summarize the Redfield QME here since the RC-QME builds on it. Using the Born-Markov and weak coupling approximations, one can readily obtain the Redfield equation for the dynamics of the reduced density matrix of the system, $\hat{\rho}(t)$. The equation reads~\cite{NitzanBook}
\begin{align}
\label{eq:Redfield_eqn}
\frac{{\rm d}\rho_{mn}(t)}{{\rm d}t} &= -\ii \omega_{mn} \rho_{mn}(t)
\nonumber \\
&- \sum_{\alpha jl} \left\{ R^{\alpha}_{mj,jl}(\omega_{lj}) \rho_{ln}(t) + [R^{\alpha}_{nl,lj}(\omega_{jl})]^* \rho_{mj}(t) \right. \nonumber \\
&- \left. \left[ R^{\alpha}_{ln,mj}(\omega_{jm}) + [R^{\alpha}_{jm,nl}(\omega_{ln})]^* \right] \rho_{jl}(t) \right\}.
\end{align} 
Here, $\omega_m$ are the eigenvalues of the system's Hamiltonian, $\hat{H}_S$, and $\omega_{mn} = \omega_m - \omega_n$. The upper indices in the Redfield dissipator correspond to a given bath, and the lower indices refer to the matrix elements in the energy eigenbasis of the system's Hamiltonian. In this derivation, the initial state of the total system+baths was written as a tensor product of the system and each individual bath.  
Importantly, in Eq. (\ref{eq:Redfield_eqn}) the impacts of the different baths are additive and there is no cooperative effect between them. This is reflected in the total dissipator written as the sum of the Redfield tensor of all baths.

The Redfield tensor depends on the matrix elements of the coupling operators and the bath equilibrium correlation functions via
\begin{align}
\label{eq:Redfield_tensor}
R^{\alpha}_{mn,jl}(\omega) &= S^{\alpha}_{mn} S^{\alpha}_{jl} \int_{0}^{\infty} {\rm d}\tau e^{\ii \omega \tau} \langle \hat{B}^{\alpha}(\tau) \hat{B}^{\alpha}(0) \rangle.
\end{align}
Here, $\hat{B}^{\alpha}(\tau) = \sum_k  t_{\alpha,k}  \left( \hat{c}_{\alpha,k}^{\dagger} e^{\ii \nu_{\alpha,k} \tau} + \hat{c}_{\alpha,k} e^{-\ii \nu_{\alpha,k} \tau} \right)$ is the bath part of the interaction Hamiltonian. $\hat S^{\alpha}$ are the coupled system's operators, which in Eq. (\ref{eq:H_original}) are Pauli matrices. 
To derive this expression, one needs to assume that the baths are stationary. Taking the continuum limit with respect to the momentum degrees of freedom of the baths, and splitting the Redfield tensor into its real and imaginary parts allows us to write
\begin{align}
\label{eq:refield_tensor_decomposition}
R^{\alpha}_{mn,jl}(\omega) &= S^{\alpha}_{mn} S^{\alpha}_{jl} \left[ \Gamma^{\alpha}(\omega) + \ii \delta^{\alpha}(\omega) \right].
\end{align}
The real functions $\Gamma^{\alpha}(\omega)$ and $\delta^{\alpha}(\omega)$ are the symmetric part of the correlation function and the lamb shift, respectively. We set the lamb shift equal to zero in our considerations ~\cite{Correa_Lamb}. To compute the real part of the Redfield tensor, the spectral density function of each bath needs to be specified; we choose the Brownian function given by
\begin{align}
\label{eq:brownian_SD}
J^{\alpha}(\omega) = \frac{4\gamma_{\alpha} \Omega_{\alpha}^2 \lambda_{\alpha}^2 \omega}{(\omega^2 - \Omega_{\alpha}^2)^2 + (2\pi\gamma_{\alpha}\Omega_{\alpha}\omega)^2}.
\end{align}
Here, $\Omega_\alpha$ corresponds to the peak of the spectral function and $\gamma_\alpha$ is its width parameter. The parameter $\lambda_\alpha$ describes the system-bath coupling strength, which in the weak coupling regime has to be the smallest energy scale. As we will see in Section \ref{sec:strong_coupling}, a narrow spectral function is crucial to utilizing the RC-QME, which handles the strong coupling regime of our models. 

In terms of the spectral function, the symmetric part of the correlation function is
 \begin{align}
 \label{eq:correlation_func}
\Gamma^{\alpha}(\omega) = \begin{cases} \pi J^{\alpha}(|\omega|) n(|\omega|) & \omega < 0 \\ \pi J^{\alpha}(\omega) [n(\omega) + 1] & \omega > 0 \\ 
\lim_{\omega\to 0}\pi J^{\alpha}(\omega) n(\omega) & \omega = 0.
\end{cases}
\end{align}
Here, $n(\omega) = (e^{\beta \omega} - 1)^{-1}$ is the Bose-Einstein distribution function with the inverse temperature $\beta=1/T$ ($k_B=1$). 

\subsection{Strong coupling: The Reaction Coordinate QME}
\label{sec:strong_coupling}

Beyond the weak system-bath coupling regime, the Born-Markov approximation cannot be justified. A powerful approach to strong coupling is the idea of {\it{Markovian embedding}}. The Reaction Coordinate procedure is an example of this framework. It relies on applying a unitary transformation that extracts a degree of freedom from each bath; the extended system is assumed to weakly couple to the rest of the baths' degrees of freedom. A collective degree of freedom extracted from the bath is called a {\it reaction coordinate}, from which the transformation takes its name. After the RC mapping, Eq.~\eqref{eq:H_original} translates to~\cite{Hughes:2009RC,NazirJCP16,Nick:2021RC,anto2023effective,Luis19}
\bea &&\hat{H}^{\text{RC}} =\Delta \hat{\sigma}^{z}+\sum_{\alpha}\left[\Omega_{\alpha}\hat{a}_{\alpha}^{\dagger}\hat{a}_{\alpha}+\lambda_{\alpha}\hat{\sigma}^{\alpha}\left(\hat{a}_{\alpha}^{\dagger}+\hat{a}_{\alpha}\right)\right]
\nonumber\\
 & &+\sum_{\alpha,k}f_{\alpha,k}\left(\hat{a}_{\alpha}^{\dagger}+\hat{a}_{\alpha}\right)\left(\hat{b}_{\alpha,k}^{\dagger}+\hat{b}_{\alpha,k}\right)+\sum_{\alpha,k}\omega_{\alpha,k}\hat{b}_{\alpha,k}^{\dagger}\hat{b}_{\alpha,k},
 \nonumber\\
\label{eq:H_RC}
\eea
where for simplicity, we set $E=0$. $\{ \hat{a}_\alpha \}$ and $\{ \hat{b}_{\alpha,k} \}$ are annihilation operators of the extracted reaction coordinates and the residual bath modes, respectively. The new coupling strengths between the reaction coordinate and the residual bath are denoted by $f_{\alpha,k}$. 
The frequencies and coupling strength of the reaction coordinates are given by $\Omega_\alpha$ and $\lambda_\alpha$, respectively. For a general spectral function of the model, we find these parameters using $\lambda^2_{\alpha} = \frac{1}{\Omega}_{\alpha} \int_0^\infty {\rm{d}}\omega \omega J^{\alpha}(\omega)$ and $\Omega^2_{\alpha} = \frac{\int_0^\infty {\rm{d}}\omega \omega^3 J^{\alpha}(\omega)}{\int_0^\infty {\rm{d}}\omega  \omega J^{\alpha}(\omega)}$~\cite{Iles:2014RC}. 
For the choice of Brownian functions, $\lambda$ and $\Omega$
are the parameters of the original bath as included in Eq. (\ref{eq:brownian_SD}).
 
We refer to the combined system+reaction coordinates as the ``extended system", described by the Hamiltonian,
\begin{align}
\label{eq:h_rc_s}
\hat{H}^{\rm{RC}}_S = \hat{H}_S + \sum_{\alpha} \left[ \Omega_{\alpha} \hat{a}^{\dagger}_{\alpha} \hat{a}_{\alpha} + \lambda_{\alpha} \hat{\sigma}^{\alpha} \left( \hat{a}^{\dagger}_{\alpha} + \hat{a}_{\alpha} \right) \right].
\end{align} 
Other terms in Eq. (\ref{eq:H_RC}) correspond to the interaction between the RCs and the residual bath, and the Hamiltonians of the residual baths. The interaction of an RC with its residual bath is described by the spectral density function $J^{\alpha}_{\rm RC}(\omega) = \sum_k |f_{k,\alpha}|^2 \delta(\omega - \omega_{\alpha,k})$; there exists a general method to find this new spectral function from the original model \cite{Nazir2018}. 
Here, using the Brownian spectral function, we get an ohmic function \cite{Nick:2021RC},
\begin{align}
\label{eq:ohmic_SD}
J^{\alpha}_{\rm RC}(\omega) = \gamma_{\alpha} \omega e^{-\omega / \Lambda_{\alpha}},
\end{align}
where $\Lambda_\alpha$ is a large energy cutoff and the dimensionless width parameter $\gamma_\alpha$ of Eq. (\ref{eq:brownian_SD}) now corresponds to the residual coupling parameter. 
Assuming small $\gamma_\alpha$, the extended system weakly couples to the bath. Under the additional assumption of Markovianity for the residual bath, the Redfield QME, Eq.~\eqref{eq:Redfield_eqn}, can be used to study the dynamics and steady-state behavior of the extended model at arbitrary $\lambda_\alpha$, now an intrinsic parameter of the extended Hamiltonian of the system. 

To evaluate the Redfield tensor (\ref{eq:Redfield_tensor}), we use the interaction Hamiltonian of the residual baths and identify $\hat{S}^{\alpha} = \hat{a}^{\dagger}_{\alpha} + \hat{a}_{\alpha}$.
Expressions for the symmetric part of the
correlation function, after the mapping, are obtained from Eq. (\ref{eq:correlation_func}).

\subsection{Strong coupling: Effective Hamiltonian theory}
\label{sec:Effective-Hamiltonian-theory}

To gain a deeper understanding of the effects of noncommuting couplings to baths on the system's dynamics, we employ the effective Hamiltonian theory introduced in Ref. \cite{anto2023effective}. It was recently extended in Ref. \cite{garwola2024open} to the case of multiple baths with noncommuting coupling operators. The EFFH method allows for deriving closed-form system Hamiltonians, which are dressed with functions of the interaction parameters. These functions reveal the impact of the environments on the system.
Briefly, using the EFFH method we (i) extract reaction coordinates from each bath, and (ii) perform a polaron transformation on the reaction coordinates. This transformation weakens the coupling of the RCs to the system while imprinting coupling parameters into the system. (iii) We truncate the spectrum of each RC to its ground state. 

Performing the effective mapping of the Hamiltonian in Eq. (\ref{eq:H_original}) gives the following formal equation \cite{garwola2024open},
\bea
&&\hat{H}^{\text{eff}} =\Delta(\hat{\sigma}^{z})^{\text{eff}} + E (\hat{\sigma}^{x})^{\text{eff}}
\nonumber\\
&&+\sum_{\alpha}\sum_{k}\left[\omega_{\alpha,k}\hat{b}_{\alpha,k}^{\dagger}\hat{b}_{\alpha,k}-\frac{2\lambda_{\alpha}f_{\alpha,k}}{\Omega_{\alpha}}(\hat{\sigma}^{\alpha})^{\text{eff}}(\hat{b}_{\alpha,k}^{\dagger}+\hat{b}_{\alpha,k})\right].\nonumber\\
\label{eq:H_eff}
\eea
We omit terms containing $\hat{S}_{\alpha}^{2}$ that would normally appear in the effective Hamiltonian \cite{anto2023effective,garwola2024open}. This is because in Eq. (\ref{eq:H_original}), $\hat S^{\alpha}$ are the Pauli matrices, and the terms with $\hat{S}_{\alpha}^{2}$ provide constant energy shifts.
The effective system operators are computed using the formula \cite{garwola2024open}
\begin{equation}
\begin{aligned}(\hat{\sigma}^{\alpha})^{\text{eff}} & =\int\hat{U}_{P}(\tilde{\boldsymbol{p}})\hat{\sigma}^{\alpha}\hat{U}_{P}^{\dagger}(\tilde{\boldsymbol{p}})\prod_{\alpha}\frac{e^{-\tilde{p}_{\alpha}^{2}}}{\sqrt{\pi}}d\tilde{p}_{\alpha},\end{aligned}
\label{eq:sigma_eff}
\end{equation}
where $\tilde{p}_{\alpha}=p_{\alpha}/\sqrt{\Omega_{\alpha}}$ is the renormalized momentum and $\hat{U}_{P}(\tilde{\boldsymbol{p}})$ is the polaron transform in the momentum representation,
\begin{equation}
\begin{aligned}\hat{U}_{P}(\tilde{\boldsymbol{p}}) & =\exp\left(-i\sqrt{2}\sum_{\alpha}\tilde{p}_{\alpha}\frac{\lambda_{\alpha}}{\Omega_{\alpha}}\hat{\sigma}^{\alpha}\right).
\end{aligned}
\label{eq:U_P}
\end{equation}
In Sec. \ref{sec:dressing}, we discuss the mapping of the XYZ and XZ models. 

The EFFH method relies on two assumptions about the original baths: (i) The mapping is based on the extraction of reaction coordinates; we assume that the original spectral functions are narrow. Specifically, we use Eq. (\ref{eq:brownian_SD}) with $\gamma_{\alpha}\ll1$.
(ii) The spectrum of each RC is truncated, leaving its ground state only. This step relies on the assumption that the reaction coordinate frequency is a large energy scale in the problem.
However, previous studies showed that the EFFH method was in fact accurate and even exact for steady state at the ultra-strong coupling limit, $\lambda_{\alpha}\gg\Omega_{\alpha}$ \cite{anto2023effective,BrettEQ}.

\section{Predictions from the Effective Hamiltonian model}
\label{sec:resultsEFFH}

The dynamics under the Effective Hamiltonian can be studied analytically, to provide closed-form {\it nonperturbative} expressions for the decoherence and relaxation rates. 
To derive the decoherence rate under the EFFH theory, we first present in Sec. \ref{sec:dressing} the dressing functions that prepare the Effective Hamiltonian; technical details are delegated to Appendix \ref{sec:AppB}.
Next, turning to the decoherence rate, we present results for the XZ model in Sec. \ref{sec:decoherence}. 
Detailed derivations, as well as the analysis of the XYZ model, are included in Appendix \ref{sec:AppC}. Guided by the predictions of the EFFH method, which suggest that enhanced coupling to a dissipative bath can suppress decoherence dynamics, we validate these predictions in Section \ref{sec:Results} through RC-QME numerical simulations.

\subsection{Dressing functions}
\label{sec:dressing}

We compute the effective system operators by evaluating Eq. (\ref{eq:sigma_eff}) with (\ref{eq:U_P}) for the XYZ model, leading to
\begin{align}
    \left( \hat{\sigma}^\alpha \right)^{\mathrm{eff}}=\kappa_\alpha(\epsilon_x,\epsilon_y,\epsilon_z) \hat{\sigma}^\alpha.
    \label{eq:kappaa}
\end{align}
For details, see Appendix \ref{sec:AppB}.
Here,  $\epsilon_\alpha\equiv\lambda_\alpha/\Omega_\alpha$ and $\kappa_{\alpha}(\epsilon_x,\epsilon_y,\epsilon_z)$ is a dressing function of the operator $\hat{\sigma}^{\alpha}$; arguments of the dressing functions denote the baths to which the system is coupled. For example, in the XZ model, $\kappa_{x}(\epsilon_x,\epsilon_z)$ is the dressing function of the operator $\hat\sigma^{x}$.
It is significant to note that in our model, the mappings of the Pauli matrices return the same Pauli matrices, only multiplied by functions of the parameters $\epsilon_\alpha$. That is, the effective model does not create new internal terms in the system's Hamiltonian or new coupling terms.
As we show in Appendix \ref{sec:AppB}, in the XZ model, we obtain 
\begin{align}
    \left( \hat{\sigma}^\alpha \right)^{\mathrm{eff}}=
    \kappa_\alpha(\epsilon_x,\epsilon_z) \hat{\sigma}^\alpha.
    \label{eq:kappaxyz}
\end{align}
In our notation, since $\epsilon_y$ does not appear in the equation above, it means that it is set to zero.
While we focus here on the XZ model, mapping other combinations of system operators leads to identical results, differing only by cyclic permutations,
\begin{align}
\kappa_x(\epsilon_x, \epsilon_z) &= \kappa_y(\epsilon_x, \epsilon_y) = \kappa_z(\epsilon_y, \epsilon_z), \notag \\
\kappa_y(\epsilon_x, \epsilon_z) &= \kappa_z(\epsilon_x, \epsilon_y) = \kappa_x(\epsilon_y, \epsilon_z), \notag \\
\kappa_z(\epsilon_x, \epsilon_z) &= \kappa_x(\epsilon_x, \epsilon_y) = \kappa_y(\epsilon_y, \epsilon_z).
\end{align}
%

\begin{figure*}[t]
\fontsize{13}{10}\selectfont 
\centering
\includegraphics[width=1.0\textwidth]{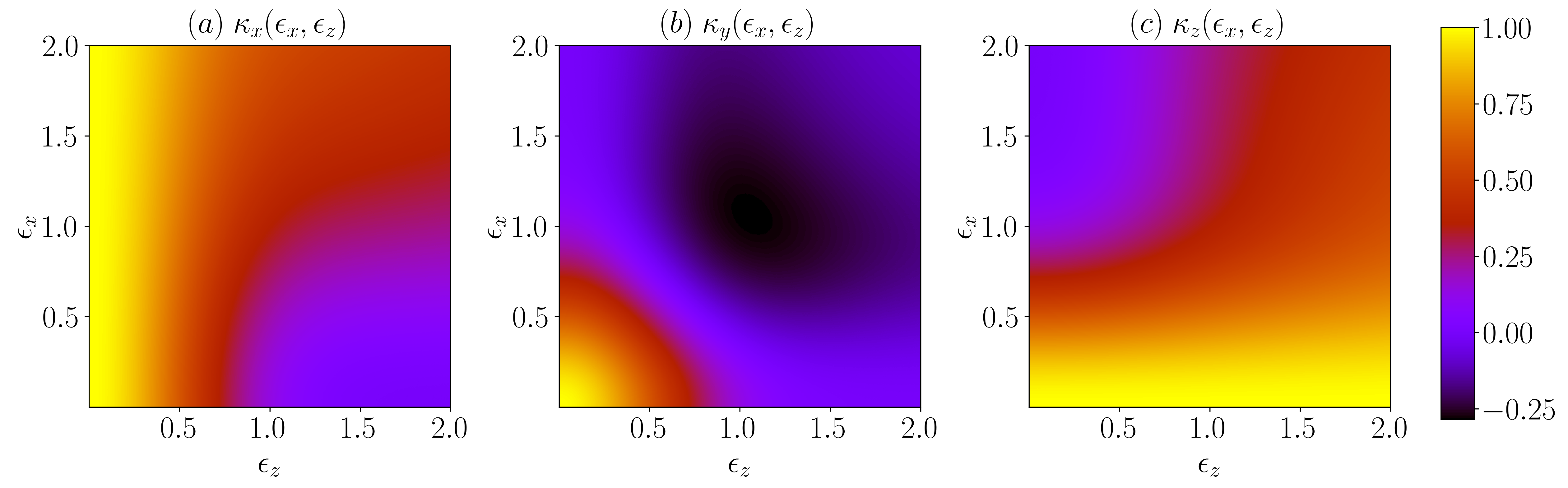}
\caption{Effective Hamiltonian theory. Presented are maps of the dressing functions for the XZ model, Eq. (\ref{eq:kappaxyz}) with (a) $\kappa_x$($\epsilon_x$, $\epsilon_z$), (b) $\kappa_y$($\epsilon_x$, $\epsilon_z$), and (c) $\kappa_z$($\epsilon_x$, $\epsilon_z$). Maps are plotted with respect to the parameters $\epsilon_\alpha=\lambda_\alpha/\Omega_\alpha$. } 
\label{fig:ZX_kappa}
\end{figure*}

\begin{figure*}[t]
\fontsize{13}{10}\selectfont 
\centering
\includegraphics[width=0.4\textwidth]{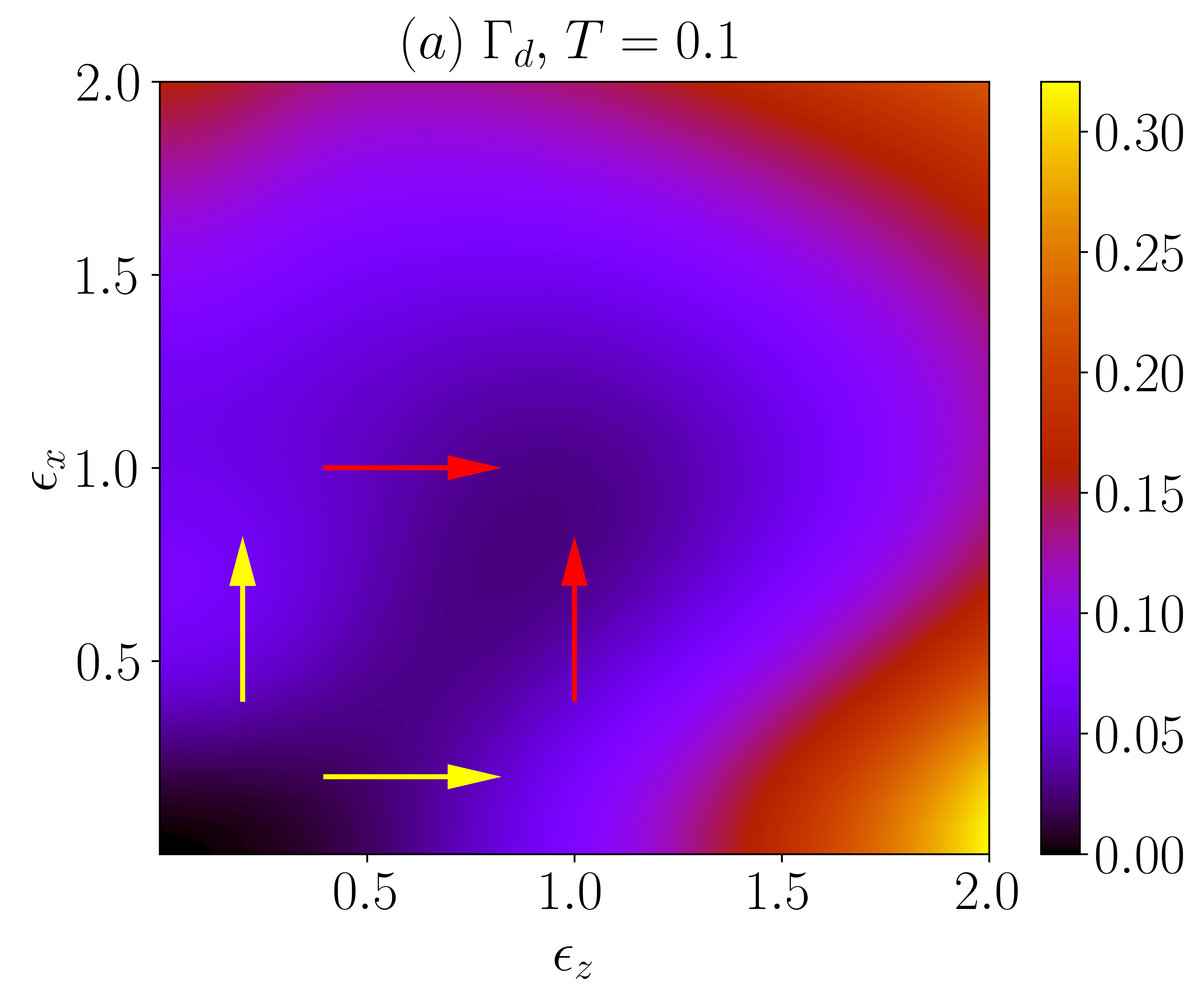}
\includegraphics[width=0.4\textwidth]{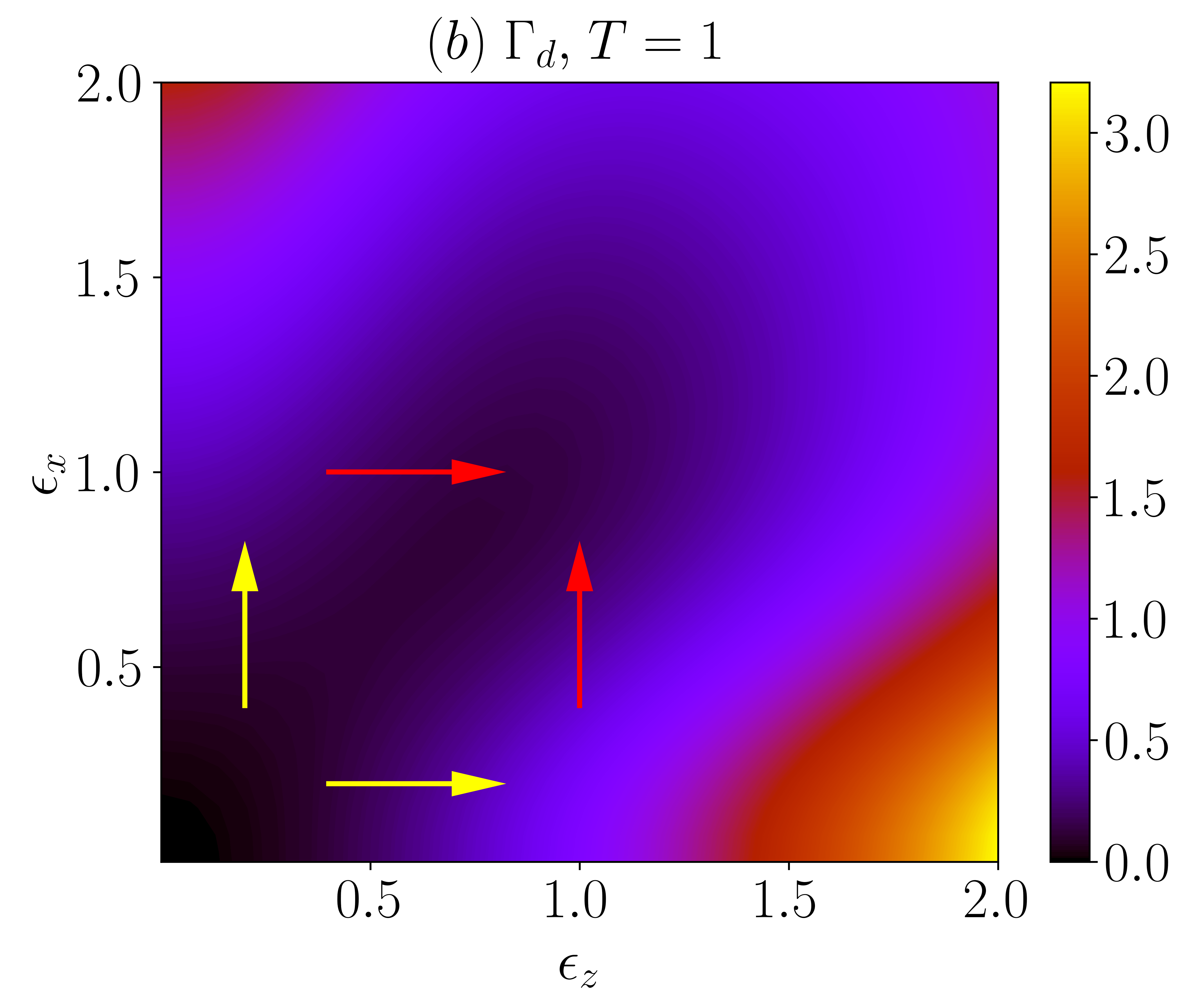}
\caption{The decoherence rate constant $\Gamma_d=\Gamma^z_\text{eff} + \Gamma^x_\text{eff}/2$,  based on
the Effective Hamiltonian theory, Eqs. 
(\ref{eq:relax_rate})-(\ref{eq:decoh_rate}),
is plotted as a function of the coupling parameters $\epsilon_z=\lambda_z/\Omega_z$ and $\epsilon_x=\lambda_x/\Omega_x$ at (a) $T = 0.1$ and (b) $T = 1$. 
Other parameters are $\Omega_z = \Omega_x = 8$,
$\gamma_z =\gamma_x$ = $\frac{0.05}{\pi}$, $\Delta$ = 1. Yellow and red arrows represent paths in the parameter space that cause the decoherence rate to increase and decrease, respectively. 
}
\label{fig:Contour}
\end{figure*}
%
\begin{figure*}[htbp]
\centering
\includegraphics[width=0.49\textwidth]{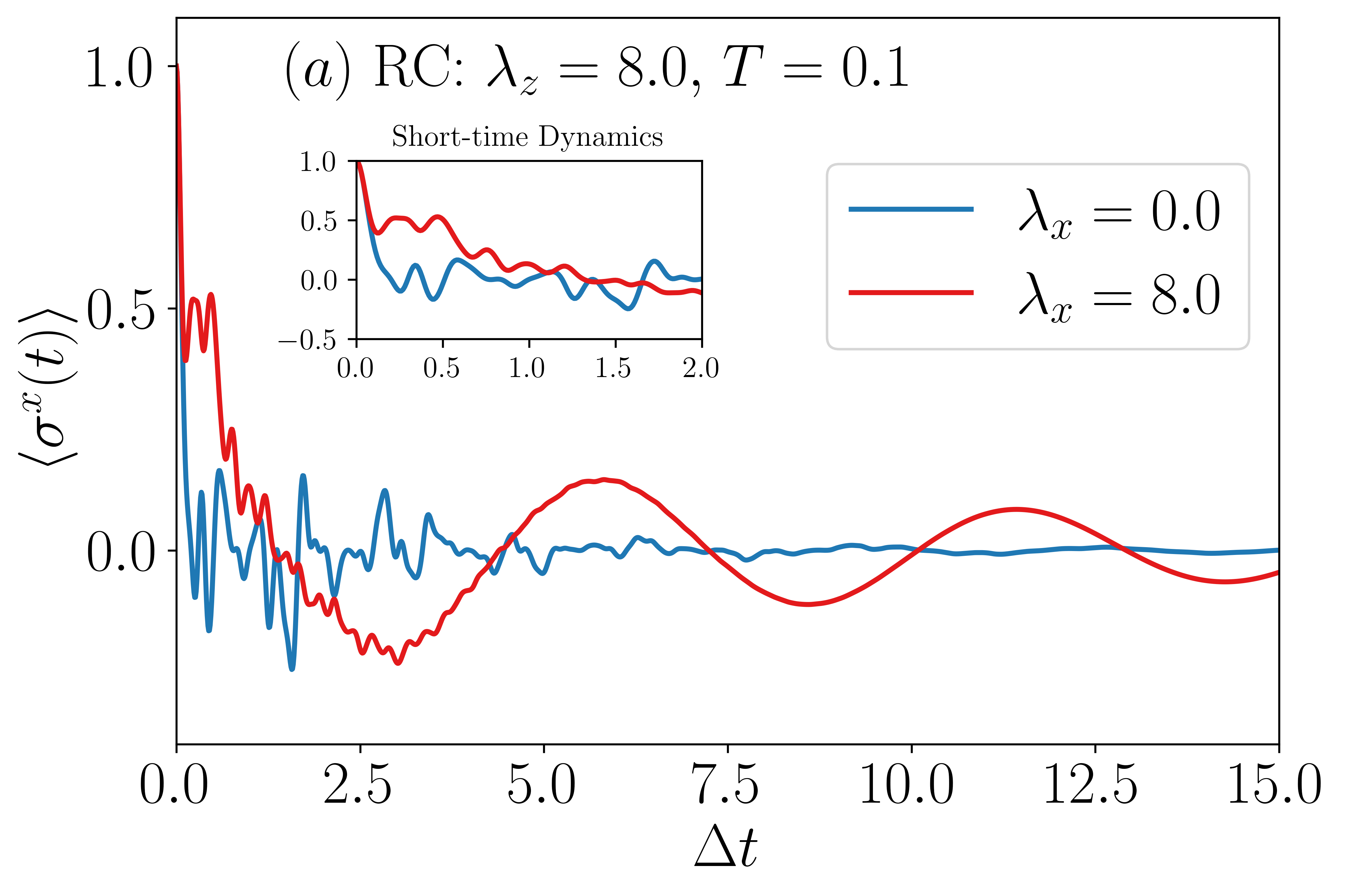}
\includegraphics[width=0.49\textwidth]{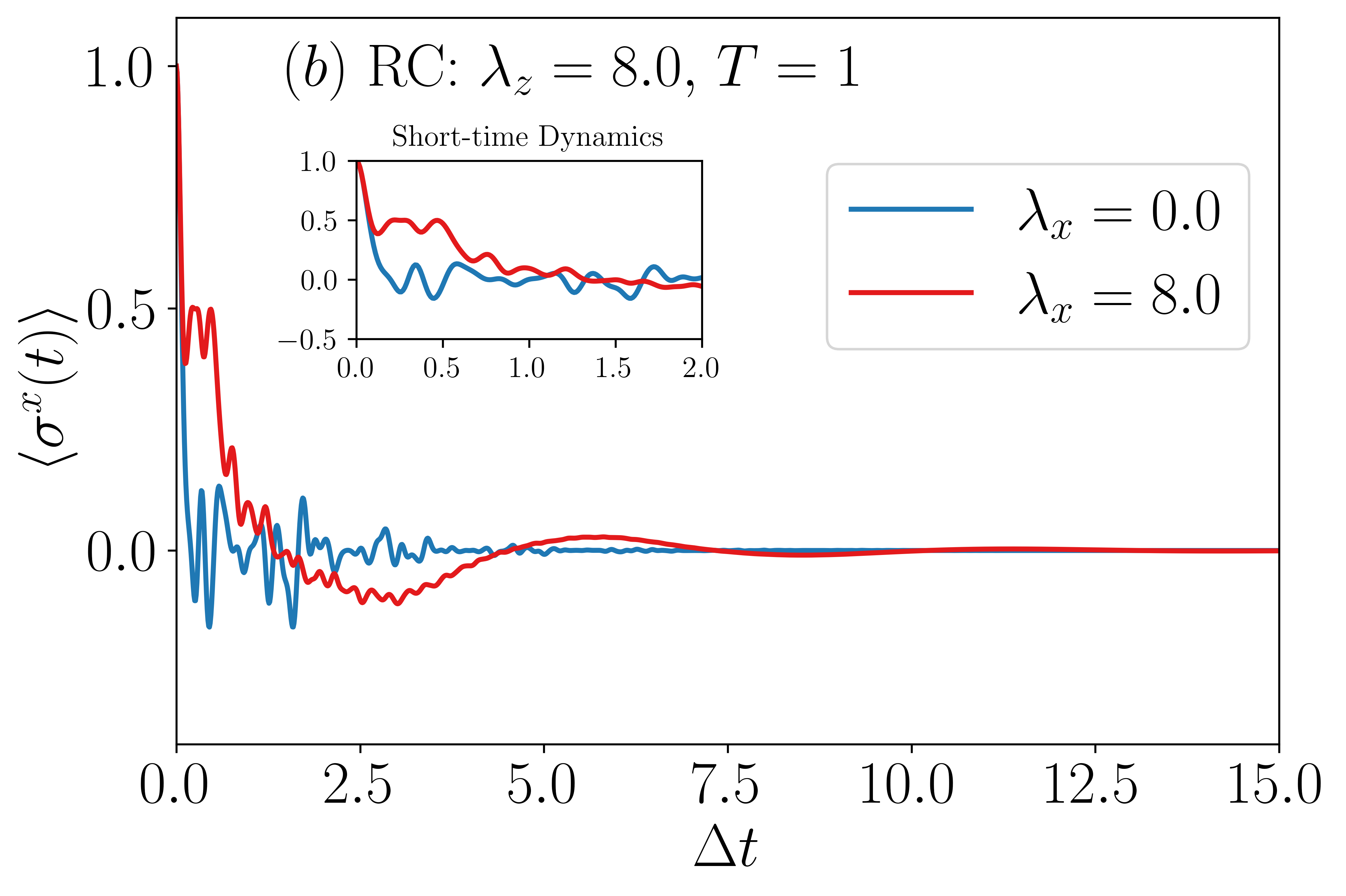}
\caption{RC-QME simulations of decoherence dynamics at temperatures (a) $T = 0.1$ and (b) $T = 1$ at strong coupling to the decohering bath, $\lambda_z=8$. We present results at $\lambda_x=0$ (blue) and $\lambda_x=8$ (red). Other parameters are $\Omega_{x,z}=8$, $\Delta =1$, $\gamma_{x,z}=0.05/\pi$. We truncate the two RCs, and use $M=5$ levels to represent each of them.
The insets focus on the short time dynamics up to $\Delta t=2$. }
\label{fig:Control_decay_rate}
\end{figure*}

\begin{figure*}[htbp]
\centering
\includegraphics[width=0.49\textwidth]{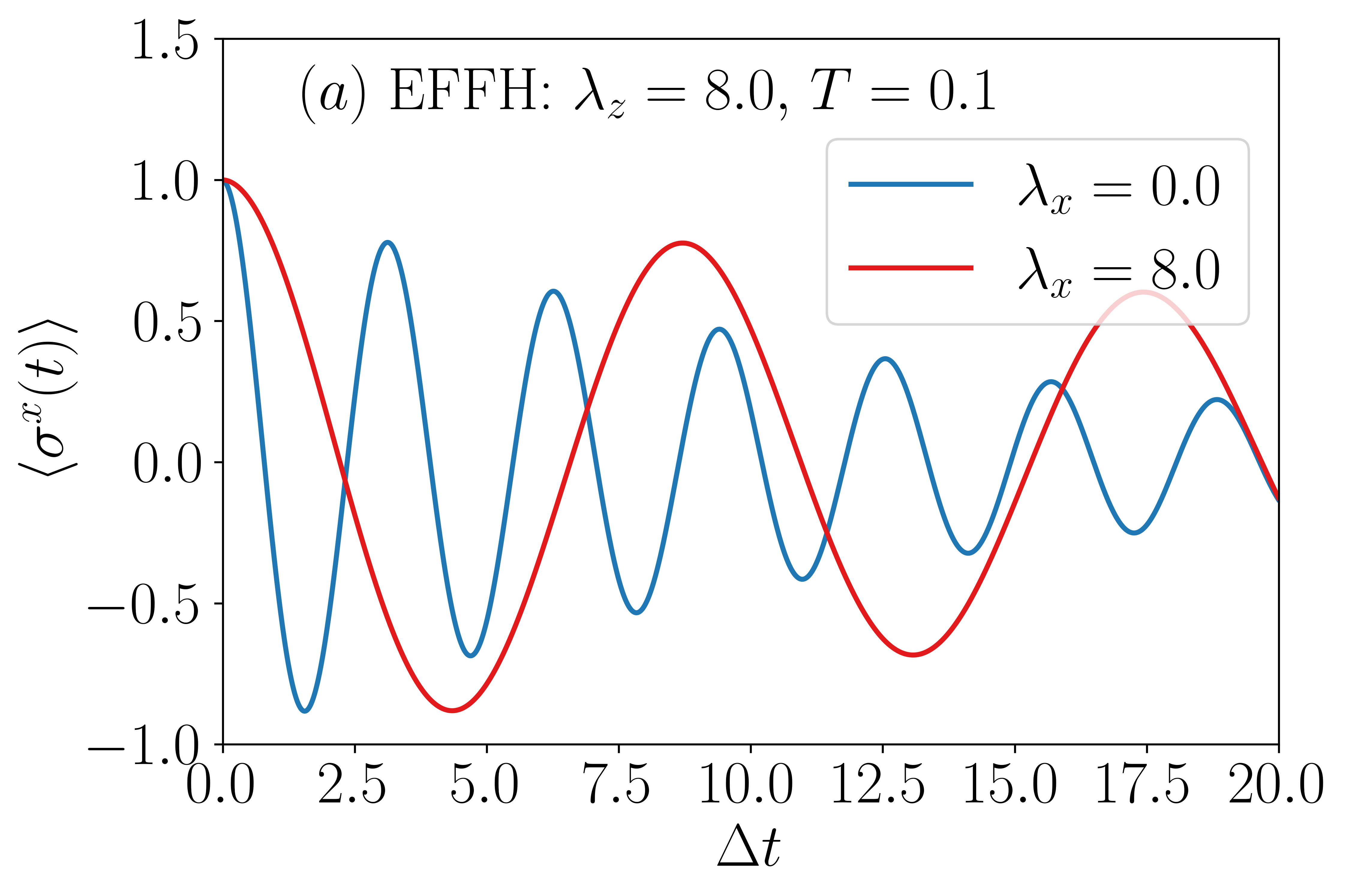}
\includegraphics[width=0.49\textwidth]{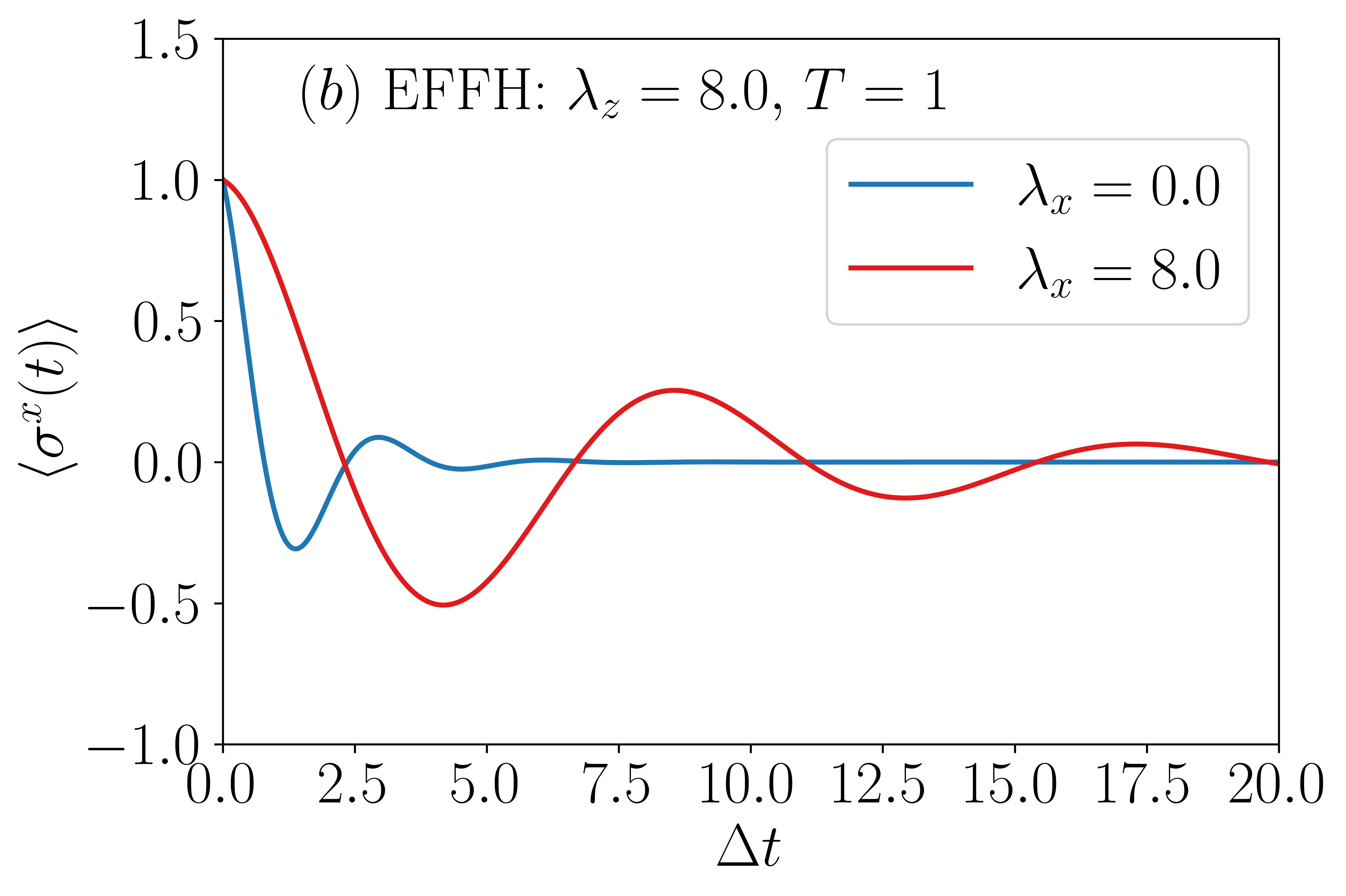}
\caption{EFFH-QME simulations of decoherence dynamics at temperatures (a) $T = 0.1$ and (b) $T = 1$ at 
strong coupling to the decoherring bath, $\lambda_z=8$. We present results at $\lambda_x=0$ (blue) and $\lambda_x=8$ (red). Other parameters are the same as in Fig. \ref{fig:Control_decay_rate}.
}
\label{fig:Control_decay_rate_EFFH}
\end{figure*}

We present the dressing functions of the XZ model in Fig. \ref{fig:ZX_kappa} with the three panels depicting (a) $\kappa_x(\epsilon_x,\epsilon_z)$,
(b) $\kappa_y(\epsilon_x,\epsilon_z)$, and (c) $\kappa_z(\epsilon_x,\epsilon_z)$. 
Beginning in panel (a), we find that
$\kappa_x$($\epsilon_x$, $\epsilon_z$) decreases monotonically as $\epsilon_z$ increases, but grows as $\epsilon_x$ increases. 
The complementary behavior is shown in panel (c): $\kappa_z$($\epsilon_x$, $\epsilon_z$) decreases when increasing $\epsilon_x$,  but increases with increasing $\epsilon_z$.
These relationships between the dressing functions will be important when we later discuss the decoherence dynamics. 
As for $\kappa_y(\epsilon_x, \epsilon_z)$ in panel (b), it reaches its upper limit 1 when both $\epsilon_z$ and $\epsilon_x$ are small (approaching zero). Interestingly, it can also become negative once both $\epsilon_z$ and $\epsilon_x$ are large. 
However, in the EFFH theory, $\kappa_y$ does not impact the decoherence dynamics of the XZ model; thus, we do not dwell on its behavior.

We now write the general form of the Effective Hamiltonian emerging from Eq. (\ref{eq:H_original}), by substituting the effective operators into Eq. (\ref{eq:H_eff}), 
\begin{widetext}
\bea
\hat{H}^{\rm eff} 
=\kappa_{z}(\epsilon_x,\epsilon_y,\epsilon_z)\Delta \hat{\sigma}^{z}+\ \kappa_{x}(\epsilon_x,\epsilon_y,\epsilon_z)E \hat{\sigma}^{x}
+
\sum_{\alpha}{\sum_{k}}\omega_{\alpha,k}\hat{b}_{\alpha,k}^{\dagger}\hat{b}_{\alpha,k}
-
\sum_{\alpha}{\sum_{k}}
2\epsilon_{\alpha}\kappa_{\alpha}(\epsilon_
x,\epsilon_y,\epsilon_z)f_{\alpha,k}\hat{\sigma}^{\alpha}
(\hat{b}_{\alpha,k}^{\dagger}+\hat{b}_{\alpha,k}).
\nonumber\\
\label{eq:HeffG}
\eea
\end{widetext}
The first two terms correspond to $\hat H_S^{\rm eff}$. They describe the mapping of the system's Hamiltonian through renormalization terms to the tunneling energy and the energy gap.
The third term describes the residual bath.
The last term corresponds to the interaction Hamiltonian between the effective system and the residual baths. 
We capture this interaction through the coupling of the $\kappa_{\alpha}(\epsilon_x,\epsilon_y,\epsilon_z)\sigma^{\alpha}$ dressed component of the system to the bosonic displacements of the baths, with the baths' spectral functions
\bea
J^{\alpha}_{\rm eff}(\omega) = \left(2\epsilon_{\alpha}\right)^2 
\sum_{k} |f_{\alpha,k}|^2\delta(\omega-\omega_{\alpha,k}).
\label{eq:JeffG}
\eea
Equation (\ref{eq:HeffG}) clearly illustrates the effects of strong coupling and noncommutativity. Due to strong coupling effects, the system's parameters become dressed as we observe in $\hat H_S^{\rm eff}$. The specific important impact of noncommutative coupling operators is that the interaction of of the system with each bath depends on its interaction with other baths, as we see in the $\kappa_{\alpha}$ prefactor in the last term of Eq. (\ref{eq:HeffG}).

\subsection{Decoherence rates}
\label{sec:decoherence}

We now focus on the XZ model in which a qubit couples to two baths through its $\hat\sigma^x$ and $\hat \sigma^z$ operators.
Considering the corresponding Effective Hamiltonian, we solve the dynamics using Redfield QME in Appendix \ref{sec:AppC}.
We find that off-diagonal elements of the reduced density matrix decay exponentially as $\exp[-\Gamma_d t]$ with
$\Gamma_d=\Gamma^{x}_{\text{eff}}/2 +  \Gamma^{z}_{\text{eff}}$ \cite{commentG}. 
The $\Gamma^{\alpha}_{\rm eff}$  rate results from the coupling of the system to the $\alpha$ bath. However, each rate reflects the impact of the other bath through (i) the renormalization of the system splitting, $\Delta \to 2\kappa_z(\epsilon_x,\epsilon_z)\Delta$ and (ii) the modification to the system bath coupling through the $\kappa_{\alpha}(\epsilon_x,\epsilon_z)$ factors. The contribution to the rate due to the $x$ bath is
\bea
\Gamma^{x}_{\text{eff}} &=&2\pi 
[\kappa_x(\epsilon_x,\epsilon_z)]^2 J^{x}_\text{eff}(2\kappa_{z}\Delta) [2n_{B}(2\kappa_{z}\Delta)+1]  
\nonumber\\ 
& \approx &2\pi[\kappa_{x}(\epsilon_x,\epsilon_z)]^{2} 
 4\epsilon_x^2 
\gamma_{x} \left(2\kappa_{z}\Delta \right)
 [2n_{B}(2\kappa_{z}\Delta)+1]
 \nonumber\\
&\xrightarrow{T\gg\kappa_z\Delta}&
16 \pi [\kappa_{x}(\epsilon_x,\epsilon_z)]^{2} 
\epsilon_x^2 \gamma_x T.
 \label{eq:relax_rate}
 \nonumber\\
\eea
The decoherence rate due to the $z$ bath is
\bea
\Gamma^{z}_{\text{eff}} & = &
\lim_{\omega\rightarrow0}2\pi J^{z}_\text{eff}(\omega)
[2n_{B}(\omega)+1] 
\nonumber\\ 
& \approx &16\pi[\kappa_{z}(\epsilon_x,\epsilon_z)]^{2}  
\epsilon_{z}^{2} 
\gamma_{z} T.
\label{eq:decoh_rate}
\eea
For clarity, in this equation and below, we adopt the short notation $\kappa_{z}$ for $\kappa_{z}(\epsilon_x,\epsilon_z)$ when dressing the energy splitting.  The approximate sign indicated that we assume large cutoff for the residual bath, $\Lambda\to \infty$.

We make the following critical observation:
The two rates, $\Gamma^{x}_{\rm eff}$ and $\Gamma_{\rm eff}^z$,
which together build the total rate of decoherence, depend each on {\it both} coupling parameters $\epsilon_z$ and $\epsilon_x$. 
Only in the weak coupling limit, $\kappa_{x,z}\to 1$ resulting in each rate depending only on the parameters of its bath. This reflects that at weak coupling, there are no cooperative effects between the baths. In contrast, at strong coupling the dressing functions introduce cooperative effects. 


In Fig. \ref{fig:Contour}, we present maps of the  decoherence rate as a function of $\epsilon_x$ and $\epsilon_z$ at two different temperatures (identical for the two baths). 
The yellow arrows depict the ``normal"  behavior:
When $\epsilon_z$ is small, indicating weak coupling to the decoherring bath, increasing this coupling, or alternatively growing the
coupling to the dissipative ($x$) bath through enhancing $\epsilon_x$, would make the decoherence rate larger.
This is the expected behavior: coupling a qubit more strongly to a decoherring bath, or to an additional dissipative environment, should accelerate the decoherence process.

However, when $\epsilon_z$ is large (e.g., $\epsilon_z=1$), we observe the counter, anomalous phenomenon. Here, as we follow the vertical red line, increasing the dissipative coupling through $\epsilon _x$ from 0 to say 1.0, {\it slows down} the decoherence dynamics. Similarly, if we couple the qubit strongly to the $x$ bath and gradually increase $\epsilon_z$, we follow the horizontal red line and observe suppression of the decoherence rate.

We can explain these normal and anomalous trends based on Eqs.
(\ref{eq:relax_rate})-(\ref{eq:decoh_rate}).
The decoherence rate $\Gamma_d$ is 
given by the sum of $ \Gamma^x_{\rm eff}/2 $ and $\Gamma^z_{\rm eff}$. When $\epsilon_z$ is small, 
$\Gamma_d$ is dominated by $\Gamma^x_{\rm eff} \propto \epsilon_x^2 \kappa_x^2$ (note that we assume high temperature in this discussion, such that we approximate $n_B(\omega)\approx T/\omega$). As we increase $\epsilon_x$, moving along the vertical yellow line, the decoherence rate increases due to the quadratic dependence $\epsilon_x^2$; the factor $\kappa_x$ does not change much with $\epsilon_x$ when $\epsilon_z$ is small, see Fig. \ref{fig:ZX_kappa}(a). This explains the normal trend of enhanced decoherence with increasing coupling to the dissipative bath.
In contrast, when $\epsilon_z$ is large, $\Gamma_d$ is dominated by $\Gamma^z_{\rm eff}\propto \epsilon_z^2 \kappa_z^2$. In this limit, an increase in $\epsilon_x$ {\it suppresses} $\kappa_z$ as we see in Fig. \ref{fig:ZX_kappa}(c), leading to an overall decrease in the decoherence rate, captured by the red vertical line.  This effect is anomalous to common expectations regarding decoherence.

The suppression of decoherence by enhanced dissipation takes place at different temperatures, as shown in Fig. \ref{fig:Contour} (a)-(b). Furthermore, based on Eqs. 
(\ref{eq:relax_rate})-(\ref{eq:decoh_rate}) and the argument above, we expect it to hold even when the baths are maintained at different temperatures.

Another interesting prediction from the EFFH theory is that we can identify scenarios in which {\it both} decoherence and relaxation rates can be simultaneously suppressed by enhanced coupling. 
As shown in Appendix \ref{sec:AppC}, the population relaxation rate is only determined by $\Gamma^x_{\rm eff}$, while the decoherence rate is given by the sum $\Gamma^x_{\rm eff}/2 + \Gamma^z_{\rm eff}$.
At large $\epsilon_x$, increasing $\epsilon_z$ suppresses $\kappa_x$, which, since $\Gamma^x_{\rm eff} \propto  \epsilon_x^2\kappa_x^2$, dominates both population relaxation and decoherence rates. Following the red horizontal line in Fig. \ref{fig:Contour} , we expect both effects to be suppressed in this regime. 

Complementing these results, in Appendix \ref{sec:AppC} we solve the dynamics of the Effective Hamiltonian using the Redfield QME.
We also introduce an additional bath coupled via the $\hat\sigma^y$ operator. We show that the decoherence rate is then determined by $(\Gamma^{x}_{\text{eff}} + \Gamma^{y}_{\text{eff}} + 2\Gamma^{z}_{\text{eff}})/2 $. Consequently, the parameters $\epsilon_x$, $\epsilon_y$, $\epsilon_z$ all influence the overall decoherence rate, and a similar cooperative effect can be expected.


\begin{figure*}[htbp]
\fontsize{13}{10}\selectfont 
\centering
\includegraphics[width=0.49\textwidth]{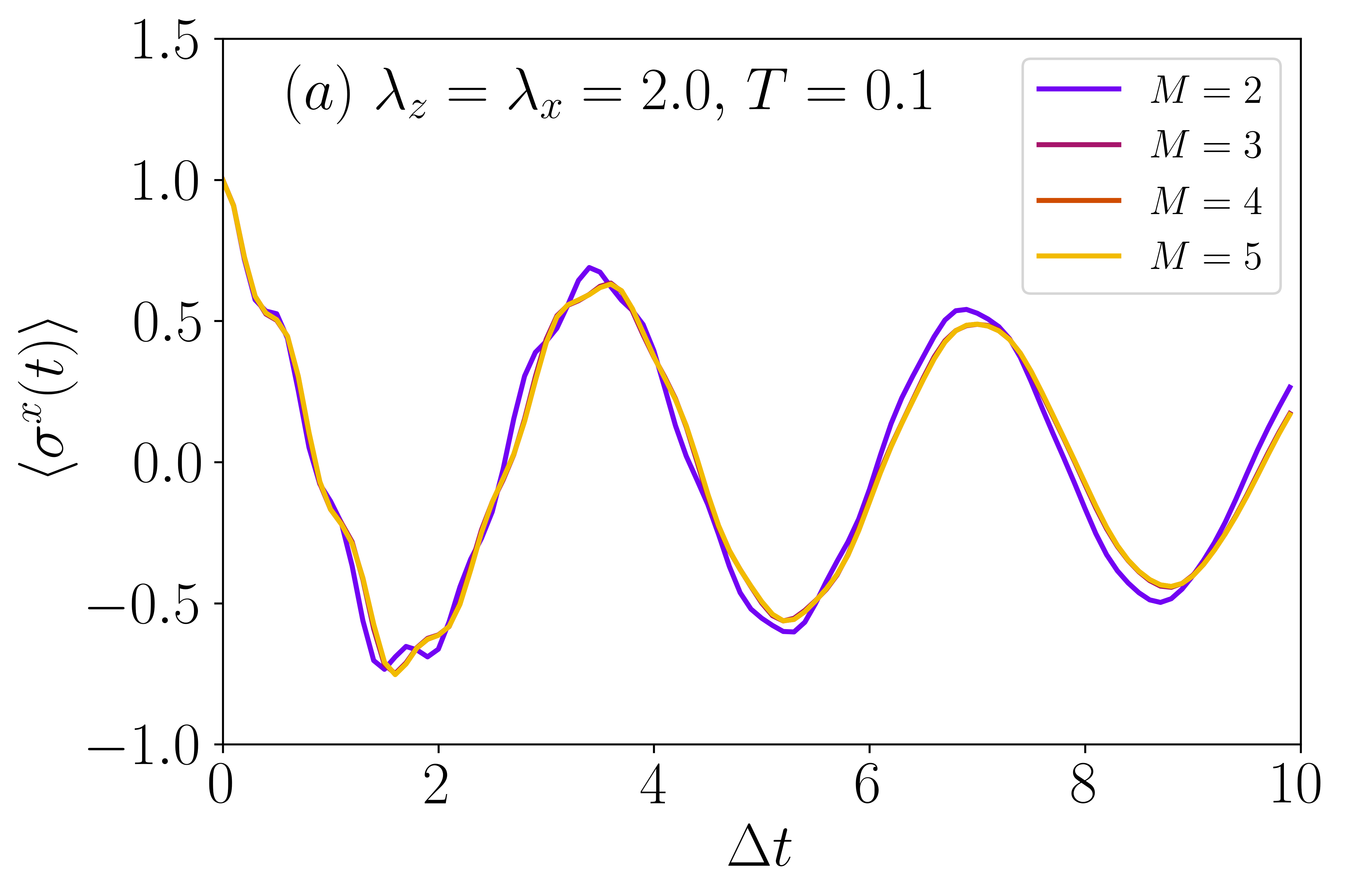}
\includegraphics[width=0.49\textwidth]{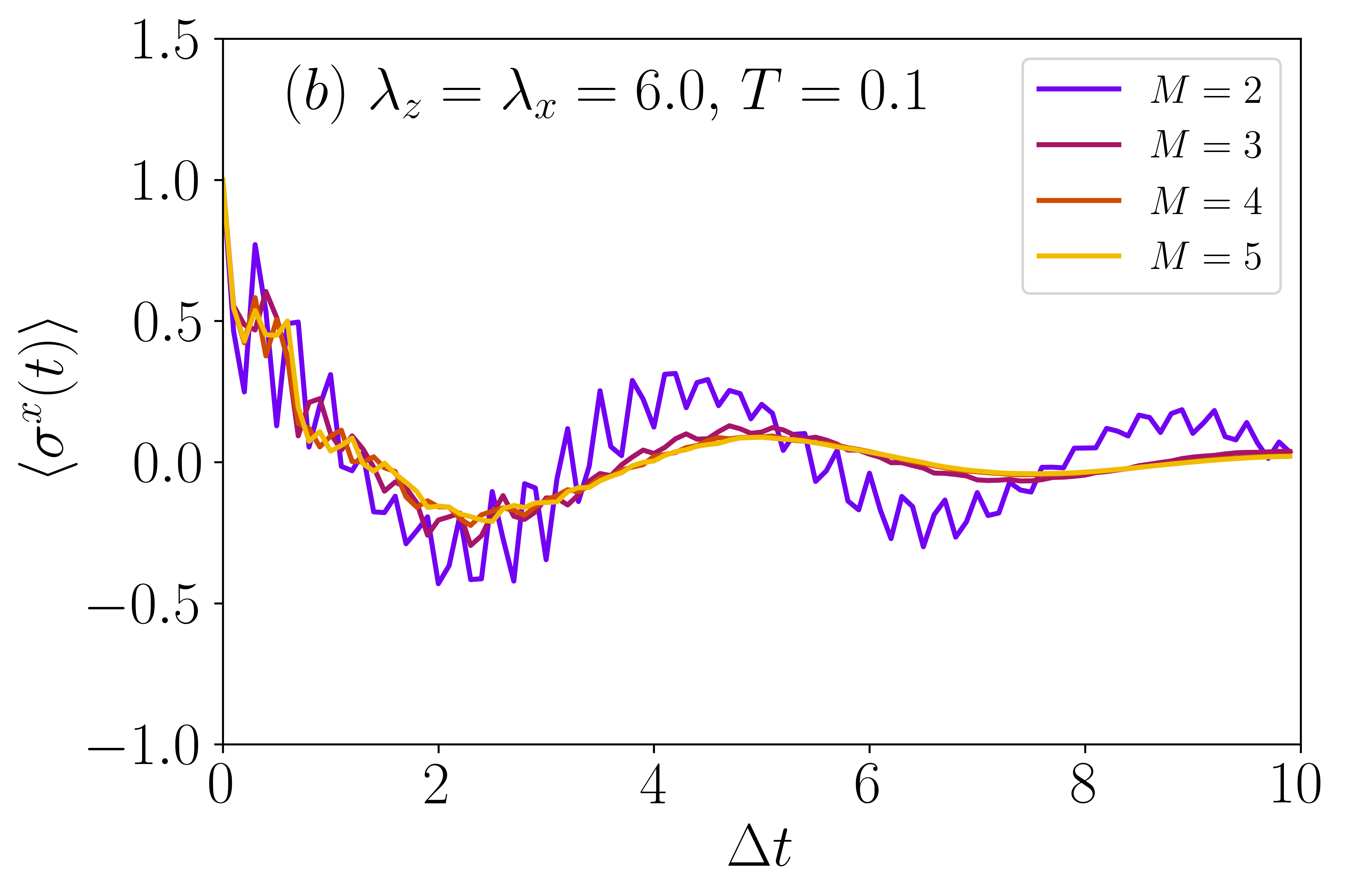}
\caption{Convergence tests for the RC-QME simulations. We present the coherences adopting different number of RC levels, $M$. We perform tests at (a) weak  and (b) strong coupling to both baths.}
\label{fig:RC_levels}
\end{figure*}

\begin{figure}[htbp]
\centering
\includegraphics[width=0.49\textwidth]{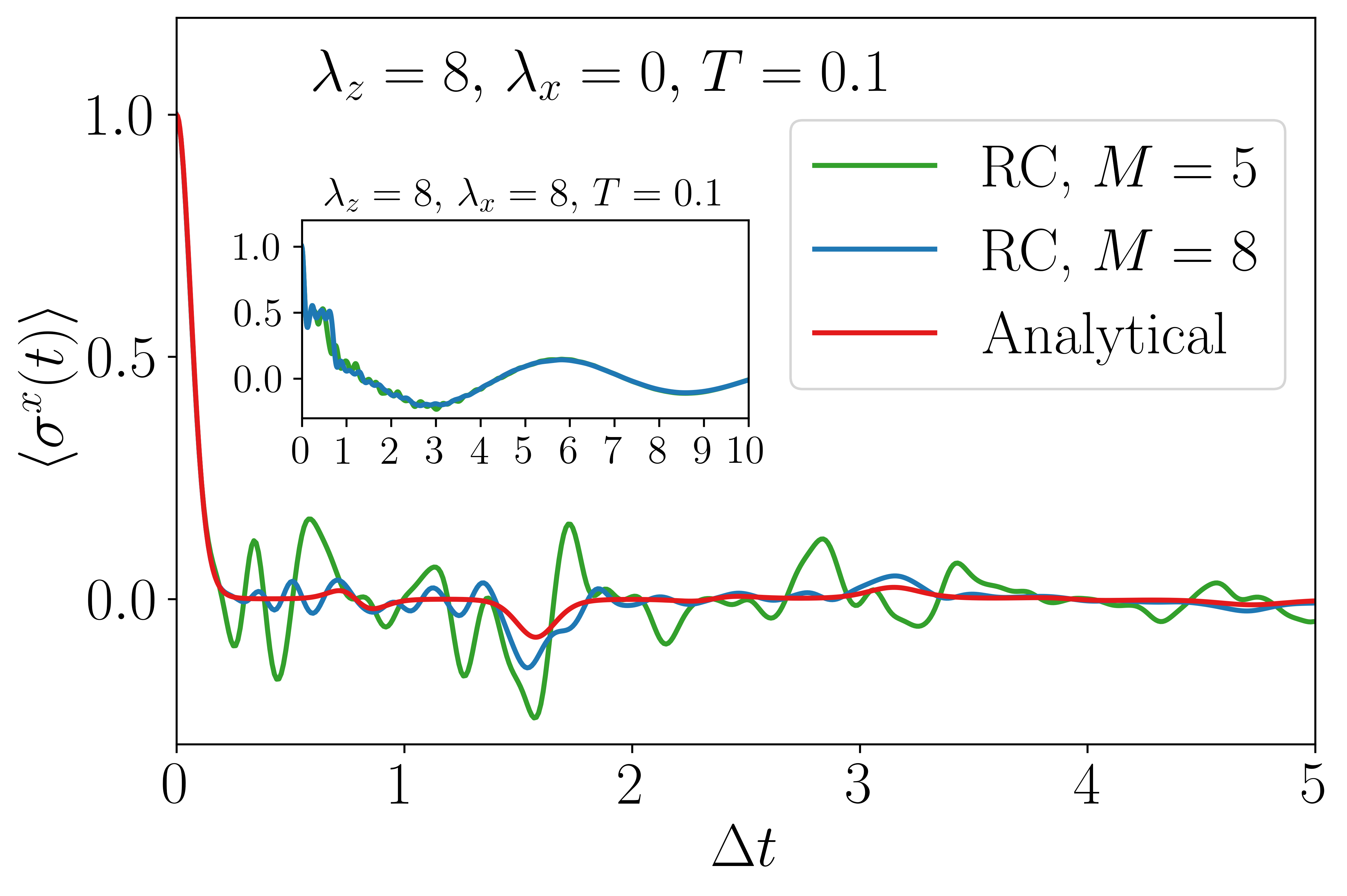} 
\caption{Convergence tests for Fig. \ref{fig:Control_decay_rate}.
Comparison of the dynamics of coherences using exact-analytical and RC-QME simulations 
 with $M=5$  and $M=8$.
The exact-analytical solution, valid at $\lambda_z=0$ is given in Eq. (\ref{eq:exactD}) \cite{BookOQS}. 
Main: Convergence test for $\lambda_x=0$, $\lambda_z=8$. Inset: 
Convergence test for $\lambda_x=8$, $\lambda_z=8$.
Other physical parameters are as in Fig. \ref{fig:Control_decay_rate} with
 $T=0.1$.
}
\label{fig:Analytical_RC_T0.1}
\end{figure}


\begin{figure*}[htpb]
\centering
\includegraphics[width=0.49\textwidth]{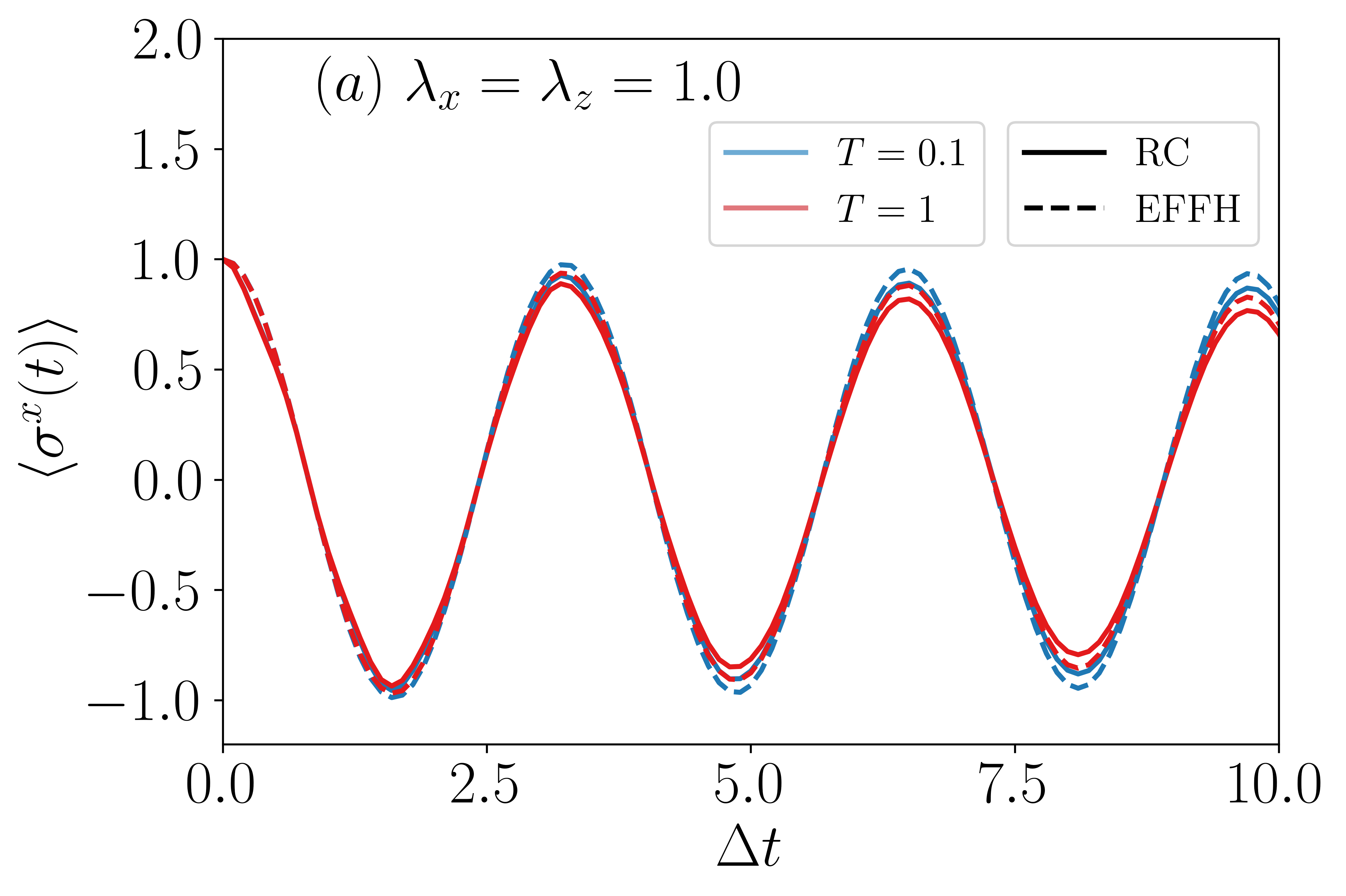}
\includegraphics[width=0.49\textwidth]{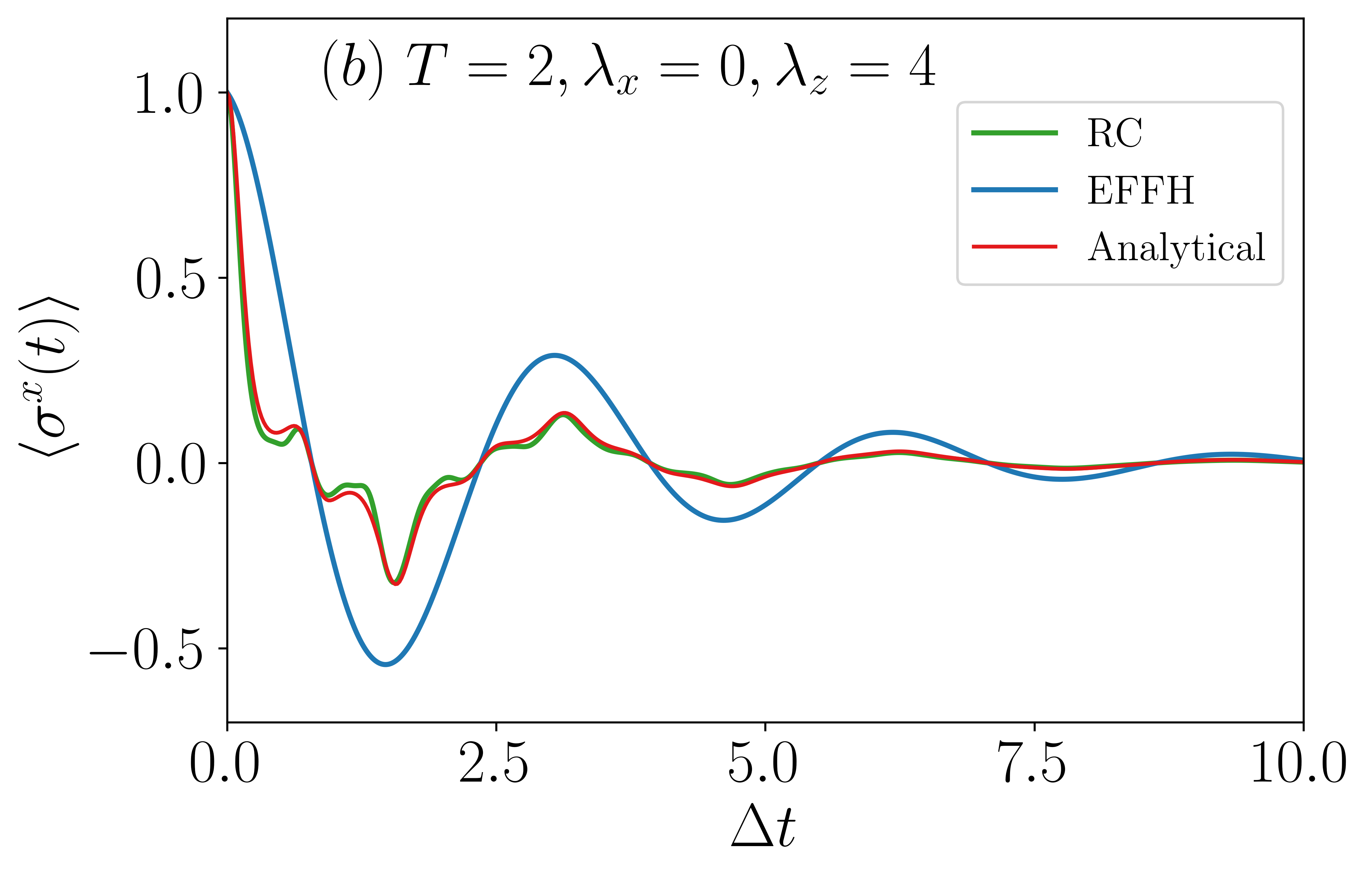}
\caption{Coherence dynamics at (a) weak and (b) intermediate coupling strength to the baths.
(a) In the weak dissipative limit, EFFH-QME simulations well agree with RC-QME results. Parameters are $T=0.1$ and $T=0.1$ with $\lambda_{x,z}=1$.
(b) In the pure decoherence model and beyond weak coupling, $\lambda_z=4$, we compare RC-QME, EFFH-QME and analytical results at $T=2$ showing qualitative correct trends of the EFFH method.}
\label{fig:EFFH_RC_weak}
\end{figure*}

\section{RC-QME simulations}
\label{sec:Results}

\subsection{Suppression of decoherence rate}
\label{subsec:Control of decoherence rate}

Guided by the predictions of the EFFH theory, we now present simulations using the RC-QME method. These results are expected to be accurate for narrow spectral functions $\gamma_{\alpha}\ll1$; a small $\gamma_{\alpha}$ implies weak system-residual bath coupling. Our chosen simulation parameters are $\Omega_z$ = $\Omega_x$ = 8, $\gamma_z$ = $\gamma_x$ = $0.05/\pi$, $\Delta$ = 1, which we keep fixed across all simulations unless otherwise stated. For the initial condition, we use $\rho_{11}=0.5$ and $\rho_{12}=0.5$ for the qubit, where 1 and 2 denote the ground state and excited state of the qubit and $\rho$ stands for the reduced density matrix of the system. The initial condition of the RC is thermal with respect to the attached bath. 
In the figures below, we present $\langle \sigma^x(t)\rangle = \rho_{12}(t)+\rho_{21}(t)$ as a function of time in dimensionless units, $\Delta t$. 

Fig. \ref{fig:Control_decay_rate} presents our main result: The suppression of decoherence by dissipation. In this example, we strongly couple the qubit to the decohering bath with $\lambda_z=8$, which, with $\Omega_{z}=8$ corresponds to $\epsilon_z=1$. In the absence of coupling to the dissipative bath, the decoherence dynamics is rapid, with coherences decaying essentially to zero at very short timescales (blue line); we discuss the subsequent rapid oscillations of the dynamics in Fig. \ref{fig:Analytical_RC_T0.1}. 
However, when we introduce the coupling to the dissipative bath, $\lambda_x \neq 0$, the decoherence process slows significantly (red). This phenomenon is illustrated at two different temperatures in Fig. \ref{fig:Control_decay_rate}(a)-(b). As expected, the decoherence process is faster at high temperatures, but similarly to the low temperature case, increasing the coupling $\lambda_x$ helps mitigating it. 

How does the EFFH method perform? 
The decoherence dynamics predicted by the EFFH method, when solved using the Redfield QME, are presented in Fig. \ref{fig:Control_decay_rate_EFFH}. Comparing these results to RC-QME results from Fig. \ref{fig:Control_decay_rate}, we find that the EFFH method is only qualitatively correct for these parameters. While the absolute values in the two figures differ, the overall trend of suppressing decoherence dynamics due to a strong $\lambda_x$ coupling remains consistent.

\subsection{Convergence tests}
\label{subsec:Reaction coordinate simulations}

To validate our main result in Fig. \ref{fig:Control_decay_rate}, we performed several convergence and benchmarking tests.
First, in Fig. \ref{fig:RC_levels}, we examine the convergence of the RC-QME simulations with respect to the truncation of the RC modes. Here, $M$ denotes the number of levels included for each RC. At weak coupling to the baths, Fig. \ref{fig:RC_levels}(a) shows that good convergence is already achieved at $M=3$. However, at strong coupling, Fig. \ref{fig:RC_levels}(b) demonstrates that more levels ($M=5$) are necessary for convergence.
Notably, this figure exhibits a ``normal" decoherence behavior: As the coupling to the dissipative bath increases, we observe a corresponding increase in the decoherence rate when comparing panels (a) and (b).

To further investigate the accuracy of the RC-QME simulations of Fig. \ref{fig:Control_decay_rate}, we analyze the coherence dynamics in a pure decoherence model in the strong coupling regime ($\lambda_z=8$, $\lambda_x=0$), comparing the RC-QME results to the exact analytical solution \cite{BookOQS}, 
\bea    \Gamma(t) &=& - 4\int_{0}^{\infty} d\omega \, J^z(\omega) \coth\left(\frac{\beta \omega}{2}\right) \frac{1 - \cos \omega t}{\omega^2}, \nonumber\\
    \langle \sigma^x (t)\rangle  &=& e^{\Gamma(t)} \cos(2\Delta t).
    \label{eq:exactD}
\eea
Here $J^z(\omega)$ is defined as in Eq.  \ref{eq:brownian_SD}. Fig. \ref{fig:Analytical_RC_T0.1} shows that as we increase $M$ from five to eight levels, the rapid oscillations up to $\Delta t=5$ are suppressed, and we get closer to the exact solution. The fast and spiky oscillations of the pure decoherence dynamics in Fig. \ref{fig:Control_decay_rate} are thus the outcome of the truncation of the RC manifold, as we also observed in Ref. \cite{antoD}.
In the inset of Fig. \ref{fig:Analytical_RC_T0.1}, we present a convergence test for the strong-coupling regime with $\lambda_{x,z}=8$. Interestingly, in this case, the convergence is already quite robust for $M=5$. Thus, while in our parameters it is challenging to converge the pure decoherence dynamics with the RC-QME, the general case with both baths active can be feasibly simulated. 

We further compare EFFH-QME to RC-QME simulations in Fig. \ref{fig:EFFH_RC_weak}. In the weak coupling limit, Fig. \ref{fig:EFFH_RC_weak}(a) shows that the EFFH method well captures the RC-QME simulations, though decoherence is minimal. As we increase the coupling energy to the moderate regime,
Fig. \ref{fig:EFFH_RC_weak}(b) demonstrates that the EFFH method captures essential features of the dynamics: the oscillation frequency and coherence lifetime. This highlights the usefulness of the EFFH predictions. Further increasing the coupling, Fig. \ref{fig:Control_decay_rate_EFFH} shows deviations from Fig. \ref{fig:Control_decay_rate} as mentioned above, yet capturing qualitative trends.

Finally, we note that we benchmarked RC-QME results against numerically exact calculations with the Hierarchical Equations of Motion in Refs. \cite{Nir24,BrettEQ}, using similar parameters as we used here. There, we confirmed the accuracy of the RC-QME simulations for both dynamics and steady state values. 
 

\section{Conclusions}
\label{sec:Conclusions}

We studied the decoherence dynamics of a qubit coupled to multiple baths through its noncommuting operators.
After presenting our methods, we first applied the EFFH framework on our model. This approach enabled us to derive closed-form analytic expressions for the decoherence rate beyond the weak coupling regime, providing insights into strategies for mitigating decoherence. The resulting decoherence rate constant (the inverse of the $T_2$ timescale) revealed how the decoherence dynamics depended on noncommutativity and coupling strengths: It demonstrated bath cooperativity and exposed conditions under which decoherence suppression may occur.

Guided by the EFFH predictions, we performed numerical simulations using the RC-QME method. When the system was weakly coupled to both the decohering and the dissipative baths, the system exhibited a normal behavior: The decoherence rate increased as either coupling strength was increased. In contrast, at strong coupling to the decoherring bath, increasing the qubit's coupling to a dissipative bath slowed down the decoherence process, defying conventional expectations. 

Our results relied on the noncommutativity of the coupling operators; the nontrivial impact of noncommuting coupling operators on quantum dynamics is gaining increasing attention across various applications. The suppression of population relaxation due to increasing interaction with an additional bath was previously discussed in Refs. \cite{NalbachPRA21, Ahsan19, garwola2024open,chen2025scalablesitespecificfrequencytuning, Archana24}. A related phenomenon: the enhancement of entanglement between a qubit and its three environments when coupled through different spin components was explored in Ref. \citenum{AndersQ24}. In Ref. \cite{feist24}, it was shown that when a system couples via noncommuting operators to both Markovian and non-Markovian baths, the presence of the Markovian environment can render the interaction with the non-Markovian bath effectively memoryless.

Coupling a system via its noncommuting operators to non-Markovian baths at strong coupling is particularly relevant in plasmonic light-matter setups. For example, this scenario applies for describing the dynamics of molecular quantum emitters in lossy nanocavities while also interacting with an intramolecular vibrational bath \cite{Ebbesen16, Ebbesen19}.
Another area of significant impact are quantum processors, where control architectures are designed to improve the $T_1$ and $T_2$ lifetimes. Our work here, alongside Ref. \citenum{garwola2024open}, suggests that frustration effects can serve as a mechanism to mitigate noise, which is supported by recent experiments \cite{Archana24}. 
A significant strength of our study lies in the analytical EFFH framework we develop, which provides guidelines on controlling open quantum system dynamics in complex settings, across a broad range of coupling regimes, multiple baths, and with noncommuting interaction operators.
Our results also have implications on the performance of qubit-based thermal conductors \cite{brenes25} and thermal machines \cite{Mu17,Friedman18,FelixQAR}, and the study of whether they can be optimized in the transient or steady-state regime with strong coupling when using noncommuting baths. 

With regard to future prospects, we hope that our study inspires experimental efforts to verify our predictions and implement this strategy for designing quantum systems with long-lived coherences.
On the theoretical side, various computational approaches, particularly numerically exact methods \cite{Kato2018,OQuPy}, could be employed to further investigate the suppression of population relaxation and decoherence dynamics through cooperative effects. 
It is also interesting to examine our model using other mapping approaches, such as the pseudo mode method \cite{Lambert19}.
Finally, our framework can be readily extended beyond single impurities to large spins and to lattice models with local couplings to multiple vibrational modes.

\begin{acknowledgments} 
D.S.~acknowledges support from an NSERC Discovery Grant. 
This research has been funded in part by the research project: ``Quantum Software Consortium: Exploring Distributed Quantum Solutions for Canada" (QSC). QSC is financed under the National Sciences and Engineering Research Council of Canada (NSERC) Alliance Consortia Quantum Grants \#ALLRP587590-23.
\end{acknowledgments}

\appendix

 \begin{widetext}
 
\section{Derivation of the effective Hamiltonian models}
\label{sec:AppB}

This Appendix is dedicated to evaluating the effective operators from the general formula Eq. (\ref{eq:sigma_eff}). We use the polaron operators (\ref{eq:U_P}) and derive the dressing functions in Eq. (\ref{eq:kappaa}) and (\ref{eq:kappaxyz}).
 For simplicity, we omit here the tilde notation for the renormalized momenta, $\tilde{p}_{\alpha}\rightarrow p_{\alpha}$. We also use the notation $\epsilon_\alpha = \lambda_\alpha/\Omega_\alpha$.

\subsection{XZ model}

We begin by deriving the expressions for the effective system operators in the XZ model. The polaron transformation is
\begin{equation}
\begin{aligned}\hat{U}_{P}(\boldsymbol{p}) & =\hat{\sigma}^{0}\cos \left(\sqrt{2} \sqrt{p_x^2 \epsilon _x^2+p_z^2 \epsilon _z^2}\right)-\frac{i \sin \left(\sqrt{2} \sqrt{p_x^2 \epsilon _x^2+p_z^2 \epsilon _z^2}\right)}{\sqrt{p_x^2 \epsilon _x^2+p_z^2 \epsilon _z^2}}\Big(\hat{\sigma}^{x}p_{x}\epsilon_{x}+\hat{\sigma}^{z}p_{z}\epsilon_{z}\Big).
\end{aligned}
\end{equation}
We obtain the effective system operators as integrals over the renormalized momenta,
\begin{equation}
\begin{aligned}(\hat{\sigma}^{x})^{\text{eff}} & =\int\frac{d\boldsymbol{p}}{\pi}\Bigg(\hat{\sigma}^{x}\frac{p_z^2 \epsilon _z^2 \cos \left(2 \sqrt{2} \sqrt{p_x^2 \epsilon _x^2+p_z^2 \epsilon _z^2}\right)+p_x^2 \epsilon _x^2}{p_x^2 \epsilon _x^2+p_z^2 \epsilon _z^2}+\hat{\sigma}^{y}\frac{p_z \epsilon _z \sin \left(2 \sqrt{2} \sqrt{p_x^2 \epsilon _x^2+p_z^2 \epsilon _z^2}\right)}{\sqrt{p_x^2 \epsilon _x^2+p_z^2 \epsilon _z^2}}\\
 & -\hat{\sigma}^{z}\frac{2 p_x p_z \epsilon _x \epsilon _z \sin ^2\left(\sqrt{2} \sqrt{p_x^2 \epsilon _x^2+p_z^2 \epsilon _z^2}\right)}{p_x^2 \epsilon _x^2+p_z^2 \epsilon _z^2}\Bigg)\\
 & =\hat{\sigma}^{x}\int\frac{d\boldsymbol{p}}{\pi}\frac{p_z^2 \epsilon _z^2 \cos \left(2 \sqrt{2} \sqrt{p_x^2 \epsilon _x^2+p_z^2 \epsilon _z^2}\right)+p_x^2 \epsilon _x^2}{p_x^2 \epsilon _x^2+p_z^2 \epsilon _z^2},
\end{aligned}
\end{equation}

\begin{equation}
\begin{aligned}(\hat{\sigma}^{y})^{\text{eff}} & =\int\frac{d\boldsymbol{p}}{\pi}\Bigg(-\hat{\sigma}^{x}\frac{p_z \epsilon _z \sin \left(2 \sqrt{2} \sqrt{p_x^2 \epsilon _x^2+p_z^2 \epsilon _z^2}\right)}{\sqrt{p_x^2 \epsilon _x^2+p_z^2 \epsilon _z^2}} +\hat{\sigma}^{y}\cos \left(2 \sqrt{2} \sqrt{p_x^2 \epsilon _x^2+p_z^2 \epsilon _z^2}\right)\\
 & +\hat{\sigma}^{z}\frac{p_x \epsilon _x \sin \left(2 \sqrt{2} \sqrt{p_x^2 \epsilon _x^2+p_z^2 \epsilon _z^2}\right)}{\sqrt{p_x^2 \epsilon _x^2+p_z^2 \epsilon _z^2}}\Bigg)\\
 & =\hat{\sigma}^{y}\int\frac{d\boldsymbol{p}}{\pi}\cos \left(2 \sqrt{2} \sqrt{p_x^2 \epsilon _x^2+p_z^2 \epsilon _z^2}\right),
\end{aligned}
\end{equation}

\begin{equation}
\begin{aligned}(\hat{\sigma}^{z})^{\text{eff}} & =\int\frac{d\boldsymbol{p}}{\pi}\Bigg(\hat{\sigma}^{x}\frac{2 p_x p_z \epsilon _x \epsilon _z \sin ^2\left(\sqrt{2} \sqrt{p_x^2 \epsilon _x^2+p_z^2 \epsilon _z^2}\right)}{p_x^2 \epsilon _x^2+p_z^2 \epsilon _z^2}-\hat{\sigma}^{y}\frac{p_x \epsilon _x \sin \left(2 \sqrt{2} \sqrt{p_x^2 \epsilon _x^2+p_z^2 \epsilon _z^2}\right)}{\sqrt{p_x^2 \epsilon _x^2+p_z^2 \epsilon _z^2}}\\
 & +\hat{\sigma}^{z}\frac{p_x^2 \epsilon _x^2 \cos \left(2 \sqrt{2} \sqrt{p_x^2 \epsilon _x^2+p_z^2 \epsilon _z^2}\right)+p_z^2 \epsilon _z^2}{p_x^2 \epsilon _x^2+p_z^2 \epsilon _z^2}\Bigg)\\
 & =\hat{\sigma}^{z}\int\frac{d\boldsymbol{p}}{\pi}\frac{p_x^2 \epsilon _x^2 \cos \left(2 \sqrt{2} \sqrt{p_x^2 \epsilon _x^2+p_z^2 \epsilon _z^2}\right)+p_z^2 \epsilon _z^2}{p_x^2 \epsilon _x^2+p_z^2 \epsilon _z^2}.
\end{aligned}
\end{equation}
Here we introduce the notation $\int\frac{d\boldsymbol{p}}{\pi}=\frac{1}{\pi}\int_{-\infty}^{\infty}dp_{x}\int_{-\infty}^{\infty}dp_{z}e^{-p_{x}^{2}-p_{z}^{2}}$. 
To compute the effective system operators, we substitute $p_{x}\rightarrow r\cos(\theta),\;p_{z}\rightarrow r\sin(\theta)$ and obtain
\bea
(\hat{\sigma}^{x})^{\text{eff}} & =&\hat{\sigma}^{x}\int_{0}^{\infty}\frac{e^{-r^{2}}r}{\pi}dr\int_{0}^{2\pi}d\theta\frac{\sin ^2(\theta ) \epsilon _z^2 \cos \left(2 \sqrt{2} r \sqrt{\cos ^2(\theta ) \epsilon _x^2+\sin ^2(\theta ) \epsilon _z^2}\right)+\cos ^2(\theta ) \epsilon _x^2}{\cos ^2(\theta ) \epsilon _x^2+\sin ^2(\theta ) \epsilon _z^2}
\nonumber\\
 & \equiv&\hat{\sigma}^{x}\kappa_{x}(\epsilon_{x},\epsilon_{z})
  =  \hat{\sigma}^{x}h(\epsilon_{x},\epsilon_{z}),
\eea
\begin{equation}
\begin{aligned}(\hat{\sigma}^{y})^{\text{eff}} & =\hat{\sigma}^{y}\int_{0}^{\infty}\frac{e^{-r^{2}}r}{\pi}dr\int_{0}^{2\pi}d\theta\cos \left(2 \sqrt{2} r \sqrt{\cos ^2(\theta ) \epsilon _x^2+\sin ^2(\theta ) \epsilon _z^2}\right)\\
 & \equiv\hat{\sigma}^{y}\kappa_{y}(\epsilon_{x},\epsilon_{z})=\hat{\sigma}^{y}f(\epsilon_{x},\epsilon_{z}),
\end{aligned}
\end{equation}

\begin{equation}
\begin{aligned}(\hat{\sigma}^{z})^{\text{eff}} & =\hat{\sigma}^{z}\int_{0}^{\infty}\frac{e^{-r^{2}}r}{\pi}dr\int_{0}^{2\pi}d\theta\frac{\cos ^2(\theta ) \epsilon _x^2 \cos \left(2 \sqrt{2} r \sqrt{\cos ^2(\theta ) \epsilon _x^2+\sin ^2(\theta ) \epsilon _z^2}\right)+\sin ^2(\theta ) \epsilon _z^2}{\cos ^2(\theta ) \epsilon _x^2+\sin ^2(\theta ) \epsilon _z^2}\\
 & \equiv\hat{\sigma}^{z}\kappa_{z}(\epsilon_{x},\epsilon_{z})=\hat{\sigma}^{z}g(\epsilon_{x},\epsilon_{z}).
\end{aligned}
\end{equation}
We introduce here three integral dressing functions of two arguments, $f$, $g$, and $h$. It turns out that three dressing functions show up for systems with other combinations of two-bath coupling operators, dressing different operators. 

In case where $\forall\alpha\;\epsilon_{\alpha}=\epsilon$ the integral functions simplify and can be expressed in terms of the Dawson integral function $F$,
\begin{equation}
\begin{aligned}h(\epsilon) & =g(\epsilon)=1-\sqrt{2}\epsilon F\left(\sqrt{2}\epsilon\right),\\
f(\epsilon) & =1-2\sqrt{2}\epsilon F\left(\sqrt{2}\epsilon\right).
\end{aligned}
\end{equation}


\subsection{XYZ model}
In the XYZ model, each bath couples to the system via the operators $\hat{\sigma}^{x}$, $\hat{\sigma}^{y}$
and $\hat{\sigma}^{z}$. The polaron transform is
\begin{equation}
\hat{U}_{P}(\boldsymbol{p})=\left(\begin{array}{cc}
\cos\left(\sqrt{2}E(\boldsymbol{p},\boldsymbol{\epsilon})\right)-\frac{ip_{z}\epsilon_{z}\sin\left(\sqrt{2}E(\boldsymbol{p},\boldsymbol{\epsilon})\right)}{E(\boldsymbol{p},\boldsymbol{\epsilon})} & \frac{\left(-p_{y}\epsilon_{y}-ip_{x}\epsilon_{x}\right)\sin\left(\sqrt{2}E(\boldsymbol{p},\boldsymbol{\epsilon})\right)}{E(\boldsymbol{p},\boldsymbol{\epsilon})}\\
\frac{\left(p_{y}\epsilon_{y}-ip_{x}\epsilon_{x}\right)\sin\left(\sqrt{2}E(\boldsymbol{p},\boldsymbol{\epsilon})\right)}{E(\boldsymbol{p},\boldsymbol{\epsilon})} & \cos\left(\sqrt{2}E(\boldsymbol{p},\boldsymbol{\epsilon})\right)+\frac{ip_{z}\epsilon_{z}\sin\left(\sqrt{2}E(\boldsymbol{p},\boldsymbol{\epsilon})\right)}{E(\boldsymbol{p},\boldsymbol{\epsilon})}
\end{array}\right),
\end{equation}
where here $E(\boldsymbol{p},\boldsymbol{\epsilon})=\sqrt{p_{x}^{2}\epsilon_{x}^{2}+p_{y}^{2}\epsilon_{y}^{2}+p_{z}^{2}\epsilon_{z}^{2}}$.
We use a coordinate transformation given by $p_{x}=r\sin(\theta)\cos(\phi),\;p_{y}=r\sin(\theta)\sin(\phi),\;p_{z}=r\cos(\theta)$
and get
\begin{equation}
\hat{U}_{P}(r,\theta,\phi)=\left(\begin{array}{cc}
\cos\left(r\sqrt{2}D(\theta,\phi,\boldsymbol{\epsilon})\right)-\frac{i\cos(\theta)\epsilon_{z}\sin\left(r\sqrt{2}D(\theta,\phi,\boldsymbol{\epsilon})\right)}{D(\theta,\phi,\boldsymbol{\epsilon})} & \frac{\sin(\theta)\left(-\epsilon_{y}\sin(\phi)-i\epsilon_{x}\cos(\phi)\right)\sin\left(r\sqrt{2}D(\theta,\phi,\boldsymbol{\epsilon})\right)}{D(\theta,\phi,\boldsymbol{\epsilon})}\\
\frac{\sin(\theta)\left(\epsilon_{y}\sin(\phi)-i\epsilon_{x}\cos(\phi)\right)\sin\left(r\sqrt{2}D(\theta,\phi,\boldsymbol{\epsilon})\right)}{D(\theta,\phi,\boldsymbol{\epsilon})} & \cos\left(r\sqrt{2}D(\theta,\phi,\boldsymbol{\epsilon})\right)+\frac{i\cos(\theta)\epsilon_{z}\sin\left(r\sqrt{2}D(\theta,\phi,\boldsymbol{\epsilon})\right)}{D(\theta,\phi,\boldsymbol{\epsilon})}
\end{array}\right),
\end{equation}
where $D(\theta,\phi,\boldsymbol{\epsilon})=\sqrt{\sin^{2}(\theta)\left(\epsilon_{x}^{2}\cos^{2}(\phi)+\epsilon_{y}^{2}\sin^{2}(\phi)\right)+\cos^{2}(\theta)\epsilon_{z}^{2}}$.
If all couplings are identical, $\epsilon_{\alpha}=\epsilon$ the polaron operator simplifies
to
\begin{equation}
\hat{U}_{P}(r,\theta,\phi)=\left(\begin{array}{cc}
\cos\left(\sqrt{2}r\epsilon\right)-i\cos(\theta)\sin\left(\sqrt{2}r\epsilon\right) & \sin(\theta)\sin\left(\sqrt{2}r\epsilon\right)(-\sin(\phi)-i\cos(\phi))\\
\sin(\theta)\sin\left(\sqrt{2}r\epsilon\right)(\sin(\phi)-i\cos(\phi)) & \cos\left(\sqrt{2}r\epsilon\right)+i\cos(\theta)\sin\left(\sqrt{2}r\epsilon\right)
\end{array}\right).
\end{equation}
In this special limit, we find that for all system operators are dressed according to
\begin{equation}
\begin{aligned}(\hat{\sigma}^{\alpha})^{\text{eff}} & =\hat{\sigma}^{\alpha}\int_{0}^{\infty}dr\int_{0}^{\pi}d\theta\int_{0}^{2\pi}d\phi\frac{r^{2}e^{-r^{2}}\sin(\theta)}{\pi^{3/2}}\left(\cos(2\theta)\sin^{2}\left(\sqrt{2}r\epsilon\right)+\cos^{2}\left(\sqrt{2}r\epsilon\right)\right)\\
 & =\hat{\sigma}^{\alpha}\left[\frac{2}{3}e^{-2\epsilon^{2}}\left(1-4\epsilon^{2}\right)+\frac{1}{3}\right].
\end{aligned}
\end{equation}
Returning to the general interaction model, the effective operators are given by the
following integral functions,
\begin{equation}
\begin{aligned}(\hat{\sigma}^{x})^{\text{eff}} & =\hat{\sigma}^{x}\int\frac{d\boldsymbol{p}}{\pi^{3/2}}\frac{\left(p_{z}^{2}\epsilon_{z}^{2}+p_{y}^{2}\epsilon_{y}^{2}\right)\cos\left(2\sqrt{2}\sqrt{p_{x}^{2}\epsilon_{x}^{2}+p_{y}^{2}\epsilon_{y}^{2}+p_{z}^{2}\epsilon_{z}^{2}}\right)+p_{x}^{2}\epsilon_{x}^{2}}{p_{x}^{2}\epsilon_{x}^{2}+p_{y}^{2}\epsilon_{y}^{2}+p_{z}^{2}\epsilon_{z}^{2}},\\
(\hat{\sigma}^{y})^{\text{eff}} & =\hat{\sigma}^{y}\int\frac{d\boldsymbol{p}}{\pi^{3/2}}\frac{\left(p_{x}^{2}\epsilon_{x}^{2}+p_{z}^{2}\epsilon_{z}^{2}\right)\cos\left(2\sqrt{2}\sqrt{p_{x}^{2}\epsilon_{x}^{2}+p_{y}^{2}\epsilon_{y}^{2}+p_{z}^{2}\epsilon_{z}^{2}}\right)+p_{y}^{2}\epsilon_{y}^{2}}{p_{x}^{2}\epsilon_{x}^{2}+p_{y}^{2}\epsilon_{y}^{2}+p_{z}^{2}\epsilon_{z}^{2}},\\
(\hat{\sigma}^{z})^{\text{eff}} & =\hat{\sigma}^{z}\int\frac{d\boldsymbol{p}}{\pi^{3/2}}\frac{\left(p_{x}^{2}\epsilon_{x}^{2}+p_{y}^{2}\epsilon_{y}^{2}\right)\cos\left(2\sqrt{2}\sqrt{p_{x}^{2}\epsilon_{x}^{2}+p_{y}^{2}\epsilon_{y}^{2}+p_{z}^{2}\epsilon_{z}^{2}}\right)+p_{z}^{2}\epsilon_{z}^{2}}{p_{x}^{2}\epsilon_{x}^{2}+p_{y}^{2}\epsilon_{y}^{2}+p_{z}^{2}\epsilon_{z}^{2}},
\end{aligned}
\end{equation}
where we introduce the short notation $\int\frac{d\boldsymbol{p}}{\pi^{3/2}}=\frac{1}{\pi^{3/2}}\int_{-\infty}^{\infty}dp_{x}\int_{-\infty}^{\infty}dp_{y}\int_{-\infty}^{\infty}dp_{z}e^{-p_{x}^{2}-p_{y}^{2}-p_{z}^{2}}$.
We already omitted terms that are odd functions of the momenta since they do will not contribute
to the integrals. In spherical coordinates
we get
\bea
(\hat{\sigma}^{x})^{\text{eff}} & =&\hat{\sigma}^{x}\int_{0}^{\infty}r^{2}dr\int_{0}^{\pi}\sin(\theta)d\theta\int_{0}^{2\pi}d\phi\frac{e^{-r^{2}}}{\pi^{3/2}}I(\theta,\phi,\epsilon_{z},\epsilon_{y},\epsilon_{x})
\nonumber\\
&=&\hat{\sigma}^{x}\kappa_{x}(\epsilon_{x},\epsilon_{y},\epsilon_{z}),
\nonumber\\
(\hat{\sigma}^{y})^{\text{eff}} & =&\hat{\sigma}^{y}\int_{0}^{\infty}r^{2}dr\int_{0}^{\pi}\sin(\theta)d\theta\int_{0}^{2\pi}d\phi\frac{e^{-r^{2}}}{\pi^{3/2}}I(\theta,\phi,\epsilon_{x},\epsilon_{z},\epsilon_{y})
\nonumber\\
&=&\hat{\sigma}^{y}\kappa_{y}(\epsilon_{x},\epsilon_{y},\epsilon_{z}),
\nonumber\\
(\hat{\sigma}^{z})^{\text{eff}} & =&\hat{\sigma}^{z}\int_{0}^{\infty}r^{2}dr\int_{0}^{\pi}\sin(\theta)d\theta\int_{0}^{2\pi}d\phi\frac{e^{-r^{2}}}{\pi^{3/2}}I(\theta,\phi,\epsilon_{y},\epsilon_{x},\epsilon_{z})
\nonumber\\
&=&\hat{\sigma}^{z}\kappa_{z}(\epsilon_{x},\epsilon_{y},\epsilon_{z}),
\eea
where
\bea
&&I(\theta,\phi,x,y,z)=
\nonumber\\
&&\frac{\sin^{2}(\theta)\left(x^{2}\cos^{2}(\phi)+y^{2}\sin^{2}(\phi)\right)\cos\left(2r\sqrt{2\sin^{2}(\theta)\left(x^{2}\cos^{2}(\phi)+y^{2}\sin^{2}(\phi)\right)+2\cos^{2}(\theta)z^{2}}\right)+\cos^{2}(\theta)z^{2}}{\sin^{2}(\theta)\left(x^{2}\cos^{2}(\phi)+y^{2}\sin^{2}(\phi)\right)+\cos^{2}(\theta)z^{2}}.
\nonumber\\
\eea
To express the momenta integrals in spherical coordinates we used a different coordinate transformation for each operator, namely
\begin{equation}
    \begin{aligned}
        \hat\sigma^x : p_y\to r \sin (\theta ) \cos (\phi ),\,\,\,\,p_z\to r \sin (\theta ) \sin (\phi ),\,\,\,\,p_x\to r \cos (\theta ),  \\
        \hat\sigma^y :p_x\to r \sin (\theta ) \cos (\phi ),\,\,\,\,p_z\to r \sin (\theta ) \sin (\phi ),\,\,\,\,p_y\to r \cos (\theta ), \\
        \hat\sigma^z :p_x\to r \sin (\theta ) \cos (\phi ),\,\,\,\,p_y\to r \sin (\theta ) \sin (\phi ),\,\,\,\,p_z\to r \cos (\theta ).
    \end{aligned}
\end{equation}

In  Fig. \ref{fig:XYZ_kappa_z_2D},
we present cuts of the dressing functions when the system is coupled to all three baths, each with a different Pauli matrix. Given the three-dimensional nature of the plots, which depend on $\epsilon_x, \epsilon_y, \epsilon_z$, we provide in
 Fig. \ref{fig:XYZ_kappa_z_2D}(a)
a two-dimensional surface plot representing the intersection of the full plot at $\epsilon_x = \epsilon_z$. Additionally,  in  Fig. \ref{fig:XYZ_kappa_z_2D}(b),
we present a series of curves showing $\kappa_z(\epsilon_x, \epsilon_y, \epsilon_z)$ values as $\epsilon_x = \epsilon_z$ varies for different $\epsilon_y$ values. 
We find that for small $\epsilon_y$, $\kappa_z$ initially decreases and then increases slightly as $\epsilon_x = \epsilon_z$ increases. In contrast, at large $\epsilon_y$, $\kappa_z$ increases monotonically with increasing $\epsilon_x = \epsilon_z$. In addition,  at small $\epsilon_x = \epsilon_z$, $\kappa_z$ vary significantly when increasing $\epsilon_y$. However, at large $\epsilon_x = \epsilon_z$, $\kappa_z$ approaches a constant value, independent of $\epsilon_y$.

\begin{figure*}[t]
\fontsize{13}{10}\selectfont 
\centering
\includegraphics[width=0.49\textwidth]{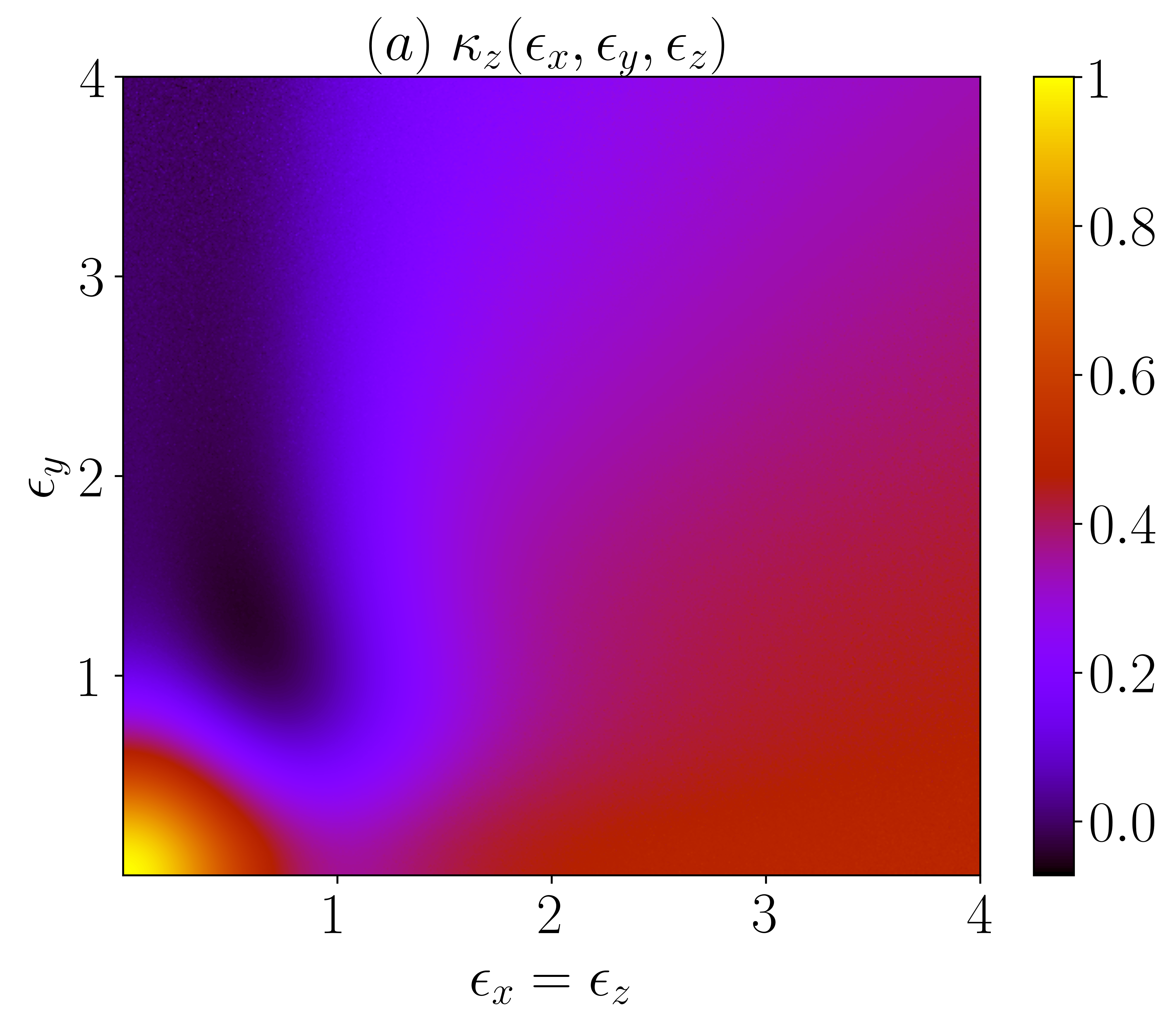}
\includegraphics[width=0.49\textwidth]{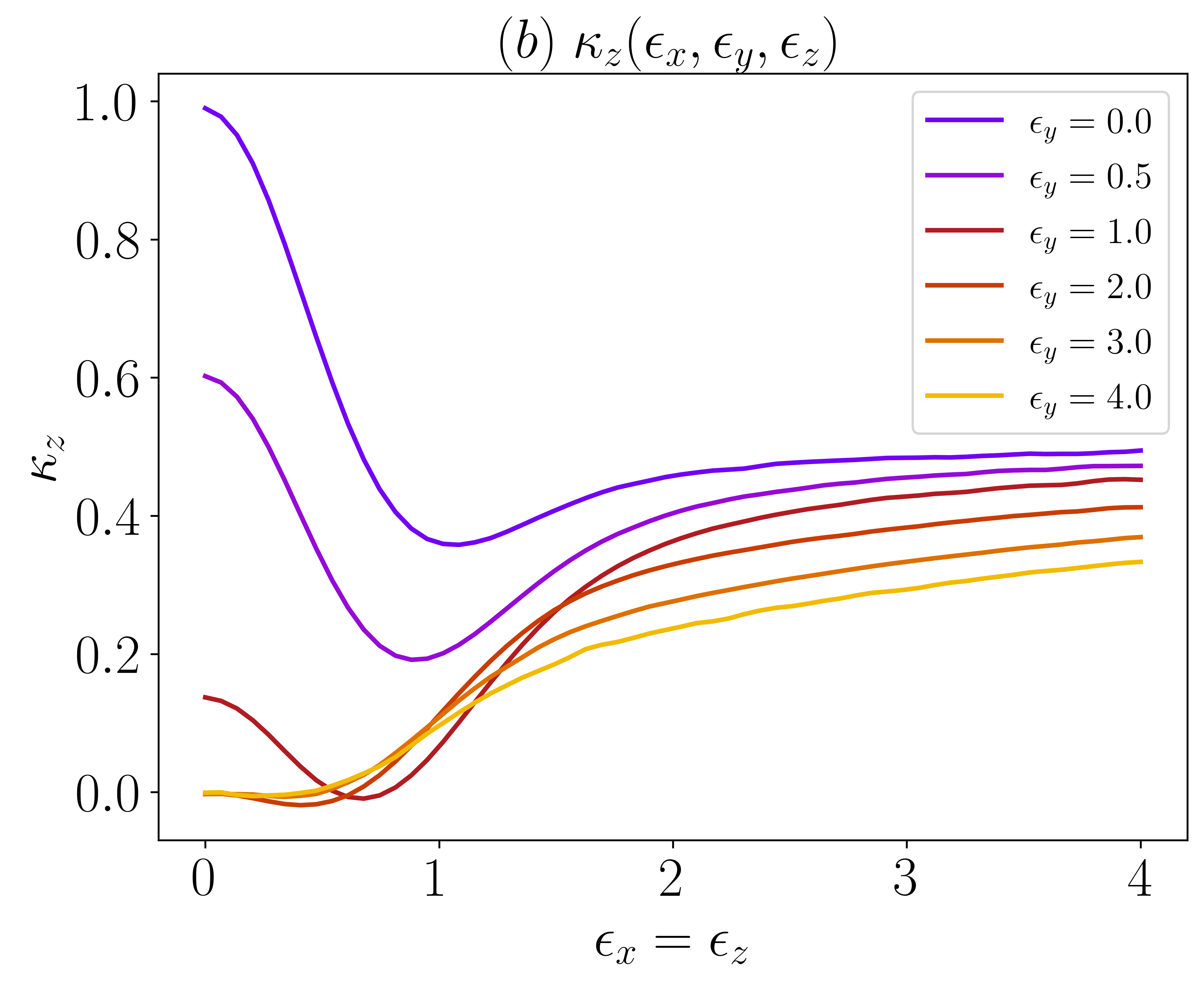}
\caption{(a) Surface plot of the function dressing the $\sigma^z$ operator in the XYZ model for $\epsilon_x=\epsilon_z$. (b) Examples  of $\kappa_z(\epsilon_x,\epsilon_y,\epsilon_z)$ with respect to $\epsilon_x = \epsilon_z$ for different values of $\epsilon_y$.}
\label{fig:XYZ_kappa_z_2D}
\end{figure*}

\section{Dynamics under the Effective Hamiltonian model using the Redfield QME}
\label{sec:AppC}

\vspace{10mm}
In this Appendix, we solve the Redfield equation for the reduced density matrix for Effective Hamiltonian models introduced in Section \ref{sec:resultsEFFH}. 

\subsection{XZ model}
\label{sec:Redfield_XZ}

Applying the general Redfield equation Eq. (\ref{eq:Redfield_eqn}) together with Eq. (\ref{eq:refield_tensor_decomposition}) and Eq. (\ref{eq:correlation_func}) on the Effective Hamiltonian, Eq. (\ref{eq:HeffG}), and reducing it to the XZ model, we get equations of motion for the coherences and population.

\subsubsection{Coherences} 

The dynamics of coherences is given by
\begin{equation}
    \begin{aligned}\dot{{\rho}}_{12}(t)= & -i\omega_{12}{\rho}_{12}(t)\\
 & +\left[-R_{11,11}^{z}(\omega_{11})-R_{22,22}^{z,*}(\omega_{22})+R_{22,11}^{z}(\omega_{11})+R_{11,22}^{z,*}(\omega_{22})\right]{\rho}_{12}(t)\\
 & -\left[R_{12,21}^{x}(\omega_{12})+R_{21,12}^{x,*}(\omega_{21})\right]{\rho}_{12}(t)+\left[R_{12,12}^{x}(\omega_{21})+R_{21,21}^{x,*}(\omega_{12})\right]{\rho}_{21}(t)\\
= & -i\omega_{12}{\rho}_{12}(t)\\
&-\left[2\pi\left(\kappa_{z}^{xz}\right)^{2}J^{z}_{\text{eff}}(0)(2n_{B}(0)+1)\right]{\rho}_{12}(t)\\
 & -\left[\pi\left(\kappa_{x}^{xz}\right)^{2}J^{x}_{\text{eff}}(\omega_{21})(2n_{B}(\omega_{21})+1)\right]{\rho}_{12}(t)+\left[\pi\left(\kappa_{x}^{xz}\right)^{2}J^{x}_{\text{eff}}(\omega_{21})(2n_{B}(\omega_{21})+1)\right]{\rho}_{21}(t).
\end{aligned}
\end{equation}
We use the compact notation $\kappa_{\alpha}^{xz}=\kappa_\alpha(\epsilon_x,\epsilon_z)$ with the baths to which the system is coupled  indicated as superscripts of $\kappa$. We also denote the qubit levels by 1 (ground) and 2 (excited), thus $\omega_{21}=-\omega_{12}=2\kappa_{z}^{xz}\Delta$.
We now identify two rate constants,
\bea
\Gamma^{x}_{\text{eff}} 
& =&2\pi\left(\kappa_{x}^{xz}\right)^{2}J^{x}_{\text{eff}}(2\kappa_{z}^{xz}\Delta)(2n_{B}(2\kappa_{z}^{xz}\Delta)+1)
\nonumber\\
&\approx&
2\pi\left(\kappa_{x}^{xz}\right)^{2}(4\epsilon_{x}^{2}\gamma_{x})\times(2\kappa_{z}^{xz}\Delta)(2n_{B}(2\kappa_{z}^{xz}\Delta)+1)
\nonumber\\
&\xrightarrow{T\gg\kappa_z^{xz}\Delta}&
16\pi \left(\kappa_{x}^{xz}\right)^{2} \epsilon_{x}^2 \gamma_x T,
\label{eq:BGax}
\eea
and
\bea
\nonumber\\
\Gamma^{z}_{\text{eff}} 
&=&\lim_{\omega\rightarrow0}2\pi\left(\kappa_{z}^{xz}\right)^{2}J^{z}_{\text{eff}}(\omega)(2n_{B}(\omega)+1)
\nonumber\\
&\approx& 2\pi\left(\kappa_{z}^{xz}\right)^{2}4\epsilon_{z}^{2}\frac{2\gamma_{z}}{\beta}
\nonumber\\
&=& 16  \pi\left(\kappa_{z}^{xz}\right)^{2}\epsilon_{z}^{2} \gamma_{z} T.
\label{eq:BGaz}
\eea
We use the approximate sign as we work in the high cutoff limit of Eq. (\ref{eq:ohmic_SD}). 
This allows us to write the equations of motion in a compact form,
\begin{equation}
    \begin{aligned}\dot{{\rho}}_{12}(t) & =-i\omega_{12}{\rho}_{12}(t)-\Gamma^{z}_{\text{eff}}{\rho}_{12}(t)-\frac{1}{2}\Gamma^{x}_{\text{eff}}{\rho}_{12}(t)+\frac{1}{2}\Gamma^{x}_{\text{eff}}{\rho}_{21}(t),\\
\dot{{\rho}}_{21}(t) & =i\omega_{12}{\rho}_{21}(t)-\Gamma^{z}_{\text{eff}}{\rho}_{21}(t)-\frac{1}{2}\Gamma^{x}_{\text{eff}}{\rho}_{21}(t)+\frac{1}{2}\Gamma^{x}_{\text{eff}}{\rho}_{12}(t).
\end{aligned}
\end{equation}
In a matrix form,
\begin{equation}
    \label{eq:REOM}
\begin{aligned}\begin{pmatrix}\dot{{{\rho}}}_{12}(t)+\dot{{{\rho}}}_{21}(t)\\
\dot{{{\rho}}}_{12}(t)-\dot{{{\rho}}}_{21}(t)
\end{pmatrix} & =\begin{pmatrix}-\Gamma^{z}_{\text{eff}} & i\omega_{21}\\
i\omega_{21} & -\Gamma^{z}_{\text{eff}}-\Gamma^{x}_{\text{eff}}
\end{pmatrix}\begin{pmatrix}{{\rho}}_{12}(t)+{{\rho}}_{21}(t)\\
{{\rho}}_{12}(t)-{{\rho}}_{21}(t)
\end{pmatrix}
\end{aligned}
\end{equation}
The eigenvalues of the matrix in the above equation are determined from the characteristic equation,
\begin{equation}
    (-\Gamma^z_{\text{eff}} - \lambda)(-\Gamma^z_{\text{eff}} - \Gamma^x_{\text{eff}} - \lambda) = (i\omega_{21})^2,
\end{equation}
leading to
\begin{equation}
    \lambda_{\pm} = \frac{- (2\Gamma^z_{\text{eff}} + \Gamma^x_{\text{eff}}) \pm \sqrt{(\Gamma^x_{\text{eff}})^2 - 4\omega_{21}^2}}{2}.
\end{equation}
The corresponding eigenvectors are
\bea
    \vec{v}_+ = 
    \begin{pmatrix}
        1 \\
        \frac{\Gamma^x_{\text{eff}} - \sqrt{(\Gamma^x_{\text{eff}})^2 - 4\omega_{21}^2}}{2\omega_{21}} \cdot i 
    \end{pmatrix}    ,\,\,\,\,\,\,\,\,\,
\vec{v}_- = 
    \begin{pmatrix}
        1 \\
        \frac{\Gamma^x_{\text{eff}} + \sqrt{(\Gamma^x_{\text{eff}})^2 - 4\omega_{21}^2}}{2\omega_{21}} \cdot i
    \end{pmatrix}
\eea

The solution to (\ref{eq:REOM}) is
\begin{equation}
    \begin{pmatrix}
        {\rho}_{12}(t) + {\rho}_{21}(t) \\
        {\rho}_{12}(t) - {\rho}_{21}(t)
    \end{pmatrix}
    =
    c_+ e^{\lambda_+ t} 
    \begin{pmatrix}
        1 \\
        \frac{\Gamma^x_{\text{eff}} - \sqrt{(\Gamma^x_{\text{eff}})^2 - 4\omega_{21}^2}}{2\omega_{21}} \cdot i
    \end{pmatrix}
    +
    c_- e^{\lambda_- t}
    \begin{pmatrix}
        1 \\
        \frac{\Gamma^x_{\text{eff}} + \sqrt{(\Gamma^x_{\text{eff}})^2 - 4\omega_{21}^2}}{2\omega_{21}} \cdot i
    \end{pmatrix},
\end{equation}
with the coefficients $c_{\pm}$ determined by 
the initial conditions. We use 
\begin{equation}
    \label{eq:init}
    {\rho}(0)=\frac{1}{2}\begin{pmatrix}1 & 1\\
1 & 1
\end{pmatrix}
\end{equation}
and get
\begin{equation}
    c_+ = \frac{1}{2} + \frac{\Gamma^x_{\text{eff}}}{4i\theta}, \quad
    c_- = \frac{1}{2} - \frac{\Gamma^x_{\text{eff}}}{4i\theta}
\end{equation}
where $4\theta^{2}=-(\Gamma^{x}_{\text{eff}})^{2}+4\omega_{21}^{2}$, 
and we assume that $\theta>0$.
Putting it together, the off diagonal elements evolve as
\begin{equation}
    \begin{aligned}
        \rho_{12} (t) &= \frac{1}{2} e^{-\frac{t}{2}(\Gamma^{x}_{\text{eff}}+2\Gamma^{z}_{\text{eff}})}
        \left[
        \cos(\theta t)+\frac{\Gamma^{x}_{\text{eff}}}{2\theta}\sin(\theta t) + i\frac{\omega_{21}}{\theta}\sin(\theta t)
        \right], \\
        \rho_{21} (t) &= \frac{1}{2} e^{-\frac{t}{2}(\Gamma^{x}_{\text{eff}}+2\Gamma^{z}_{\text{eff}})}
        \left[
        \cos(\theta t)+\frac{\Gamma^{x}_{\text{eff}}}{2\theta}\sin(\theta t) - i\frac{\omega_{21}}{\theta}\sin(\theta t)
        \right].
    \end{aligned}
\end{equation}
Finally, we identify the decoherence rate by $\Gamma_d=\Gamma_{\rm eff}^x/2 + \Gamma_{\rm eff}^z$.

\subsubsection{Population}

We write now the Redfield equations of motion for the population, ${\rho}_{11}(t)$ and ${\rho}_{22}(t)$ = $1 - {\rho}_{11}(t)$. We get
\begin{equation}
    \begin{aligned} \dot{{\rho}}_{11}(t) = & -\left[ R_{12,21}^{x}(\omega_{12}) + R_{12,21}^{x,*}(\omega_{12}) \right] {\rho}_{11}(t)
  +\left[R_{21,12}^{x}(\omega_{21})+R_{21,12}^{x,*}(\omega_{21})\right]{\rho}_{22}(t)\\
= &
-\left[2\pi\left(\kappa_{x}^{xz}\right)^{2}J^{x}_{\text{eff}}(\omega_{21})n_{B}(\omega_{21})\right]{\rho}_{11}(t)
+\left[2\pi\left(\kappa_{x}^{xz}\right)^{2}J^{x}_{\text{eff}}(\omega_{21})(n_{B}(\omega_{21})+1)\right]{\rho}_{22}(t)\\
= &
-\left[2\pi\left(\kappa_{x}^{xz}\right)^{2}J^{x}_{\text{eff}}(\omega_{21})(2n_{B}(\omega_{21})+1)\right]{\rho}_{11}(t)+\left[2\pi\left(\kappa_{x}^{xz}\right)^{2}J^{x}_{\text{eff}}(\omega_{21})(n_{B}(\omega_{21})+1)\right].
\end{aligned}
\end{equation}
Adopting the definition of $\Gamma^{x}_{\text{eff}}$ of Eq. (\ref{eq:BGax}), we have
\begin{equation}
    \begin{aligned}\dot{{\rho}}_{11}(t) & =-\Gamma^{x}_{\text{eff}}{\rho}_{11}(t)+\Gamma^{x}_{\rm eff}\frac{n_{B}(\omega_{21})+1}{2n_{B}(\omega_{21})+1},\\
\dot{{\rho}}_{22}(t) & =-\Gamma^{x}_{\text{eff}}{\rho}_{22}(t)+\Gamma^{x}_{\rm eff}\frac{n_{B}(\omega_{21})}{2n_{B}(\omega_{21})+1}.
\end{aligned}
\end{equation}
Based on the initial condition, ${\rho_{11}}={\rho_{22}}=\frac{1}{2}$, the solution for the population is
\begin{equation}
\begin{aligned}
{\rho_{11}} (t)&= \left( \frac{1}{2} - \frac{e^{\beta \omega_{21}}}{e^{\beta \omega_{21}} + 1} \right) e^{-\Gamma^{x}_{\text{eff}} t} + \frac{e^{\beta \omega_{21}}}{e^{\beta \omega_{21}} + 1}, \\
{\rho_{22}} (t)&= \left( -\frac{1}{2} + \frac{e^{\beta \omega_{21}}}{e^{\beta \omega_{21}} + 1} \right) e^{-\Gamma^{x}_{\text{eff}} t} + \frac{1}{e^{\beta \omega_{21}} + 1}.
\end{aligned}
\end{equation}
Clearly, in the long time dynamics, the system thermalizes to the temperature $1/\beta$.

\subsection{XYZ model}
\label{sec:Redfield_XYZ}

Applying the general Redfield equation Eq. (\ref{eq:Redfield_eqn}) together with Eq. (\ref{eq:refield_tensor_decomposition}) and Eq. (\ref{eq:correlation_func}) on the Effective Hamiltonian, Eq. (\ref{eq:HeffG}), we get equations of motion for the coherences and population.

\subsubsection{Coherences}
The coherences evolve according to
\begin{equation}
\begin{aligned}\begin{pmatrix}\dot{{{\rho}}}_{12}+\dot{{{\rho}}}_{21}\\
\dot{{{\rho}}}_{12}-\dot{{{\rho}}}_{21}
\end{pmatrix} & =\begin{pmatrix}-\Gamma^{z}_{\text{eff}}-\Gamma^{y}_{\text{eff}}& i\omega_{21}\\
i\omega_{21} & -\Gamma^{z}_{\text{eff}}-\Gamma^{x}_{\text{eff}}
\end{pmatrix}\begin{pmatrix}{{\rho}}_{12}(t)+{{\rho}}_{21}(t)\\{{\rho}}_{12}(t)-{{\rho}}_{21}(t)
\end{pmatrix},
\end{aligned}
\end{equation}
where 
\bea \Gamma^{x}_{\text{eff}} & =&2\pi\left(\kappa_{x}^{xyz}\right)^{2}J^{x}_{\text{eff}}(2\kappa_{z}^{xyz}\Delta)(2n_{B}(2\kappa_{z}^{xyz}\Delta)+1)
\nonumber\\
&\approx&2\pi\left(\kappa_{x}^{xyz}\right)^{2}4\epsilon_{x}^{2}\gamma_{x}2\kappa_{z}^{xyz}\Delta(2n_{B}(2\kappa_{z}^{xyz}\Delta)+1)
\nonumber\\
&\xrightarrow{T \gg \kappa_z^{xyz} \Delta}&
16\pi \left(\kappa_{x}^{xyz}\right)^{2} \epsilon_{x}^2 \gamma_x T,
\nonumber\\
    \Gamma^{y}_{\text{eff}} & =&2\pi\left(\kappa_{y}^{xyz}\right)^{2}J^{y}_{\text{eff}}(2\kappa_{z}^{xyz}\Delta)(2n_{B}(2\kappa_{z}^{xyz}\Delta)+1)
    \nonumber\\
&\approx&2\pi\left(\kappa_{y}^{xyz}\right)^{2}4\epsilon_{y}^{2}\gamma_{y}2\kappa_{z}^{xyz}\Delta(2n_{B}(2\kappa_{z}^{xyz}\Delta)+1)
\nonumber\\
&\xrightarrow{T \gg \kappa_z^{xyz} \Delta}&
16\pi \left(\kappa_{y}^{xyz}\right)^{2} \epsilon_{y}^2 \gamma_y T,
    \nonumber\\
\Gamma^{z}_{\text{eff}} 
& =&\lim_{\omega\rightarrow0}2\pi\left(\kappa_{z}^{xyz}\right)^{2}J^{z}_{\text{eff}}(\omega)(2n_{B}(\omega)+1)
\nonumber\\
&\approx&2\pi\left(\kappa_{z}^{xyz}\right)^{2}4\epsilon_{z}^{2}\frac{2\gamma_{z}}{\beta}
\nonumber\\
&=& 16  \pi\left(\kappa_{z}^{xyz}\right)^{2}\epsilon_{z}^{2} \gamma_{z} T.
\eea
where $\kappa_{\alpha}^{xyz}=\kappa_\alpha(\epsilon_x,\epsilon_y,\epsilon_z)$. 
We use the approximate sign since we work in the $\Lambda$-large limit of Eq. (\ref{eq:ohmic_SD}).
Using the initial condition (\ref{eq:init}), we get
\begin{equation}
    \begin{aligned}
        \begin{pmatrix}\dot{{{\rho}}}_{12}+\dot{{{\rho}}}_{21}\\
\dot{{{\rho}}}_{12}-\dot{{{\rho}}}_{21}
\end{pmatrix}=e^{-\frac{t}{2}(\Gamma^{x}_{\text{eff}}+\Gamma^{y}_{\text{eff}}+2\Gamma^{z}_{\text{eff}})}\begin{pmatrix}\cos(\theta t)+\frac{(\Gamma^{x}_{\text{eff}}-\Gamma^{y}_{\text{eff}})}{2\theta}\sin(\theta t)\\
i\frac{\omega_{21}}{\theta}\sin(\theta t)
\end{pmatrix},
    \end{aligned}
\end{equation}
where $4\theta^{2}=-(\Gamma^{x}_{\text{eff}}-\Gamma^{y}_{\text{eff}})^{2}+4\omega_{21}^{2}$, and we assume that $\theta>0$.
Separately, the coherence dynamics is given by
\begin{equation}
\begin{aligned}
\rho_{12} (t)&= \frac{1}{2} e^{-\frac{t}{2} \left( \Gamma^{x}_{\text{eff}} + \Gamma^{y}_{\text{eff}} + 2\Gamma^{z}_{\text{eff}} \right)}
\left[ \cos (\theta t) + \frac{\Gamma^{x}_{\text{eff}} - \Gamma^{y}_{\text{eff}}}{2\theta} \sin (\theta t) + \frac{i\omega_{21}}{\theta} \sin (\theta t) \right], \\
\rho_{21} (t)&= \frac{1}{2} e^{-\frac{t}{2} \left( \Gamma^{x}_{\text{eff}} + \Gamma^{y}_{\text{eff}} + 2\Gamma^{z}_{\text{eff}} \right)}
\left[ \cos (\theta t) + \frac{\Gamma^{x}_{\text{eff}} - \Gamma^{y}_{\text{eff}}}{2\theta} \sin (\theta t) - \frac{i\omega_{21}}{\theta} \sin (\theta t) \right].
\end{aligned}
\end{equation}

\subsubsection{Population} 

For population dynamics we get
\begin{equation}
\begin{aligned}
{\rho}_{11} (t) &= \left( \frac{1}{2} - \frac{e^{\beta \omega_{21}}}{e^{\beta \omega_{21}} + 1} \right) e^{-(\Gamma^{x}_{\text{eff}}+\Gamma^{y}_{\text{eff}})t} + \frac{e^{\beta \omega_{21}}}{e^{\beta \omega_{21}} + 1}, \\
{\rho}_{22} (t) &= \left( -\frac{1}{2} + \frac{e^{\beta \omega_{21}}}{e^{\beta \omega_{21}} + 1} \right) e^{-(\Gamma^{x}_{\text{eff}}+\Gamma^{y}_{\text{eff}}) t} + \frac{1}{e^{\beta \omega_{21}} + 1}.
\end{aligned}
\end{equation}

In the XZ model, the population decay rate is proportional to $\Gamma^{x}_{\text{eff}}$. As such, stronger coupling to the decoherring bath slows down the relaxation dynamics, as $\kappa_{x}^{xz}$ is suppressed with increasing $\epsilon_z$ \cite{garwola2024open}. In the XYZ or XY models, the decay rate is proportional to $\Gamma^{x}_{\text{eff}}+ \Gamma^{y}_{\text{eff}}$, meaning that both $\kappa_{x}$ and $\kappa_{y}$ contribute. The dressing functions exhibit opposing effects, making the influence of one bath's coupling on the relaxation rate non-monotonic, with the behavior depending on the coupling strength of the other bath(s). In such cases, similar trends to the decoherence dynamics in the XZ model can be expected. As discussed in this work, when the system is strongly coupled to one bath, increasing the coupling to another bath coupled through a noncommuting operator can slow down the decay process.


\end{widetext}

\bibliography{reference}
\end{document}